\documentclass[prd, twocolumn]{revtex4}
\usepackage{graphicx}
\usepackage{amssymb}

\begin{document}

\title{{\large \bf Low Energy Branes, Effective Theory and Cosmology}}

\author{Gonzalo A. Palma and Anne-Christine Davis}

\affiliation{Department of Applied Mathematics and Theoretical
Physics, \mbox{Center for Mathematical Sciences, University of
Cambridge}, Wilberforce Road, Cambridge CB3 0WA, \\ United
Kingdom.}

\date{June 2004}

\begin{abstract}
The low energy regime of cosmological BPS-brane configurations
with a bulk scalar field is studied. We construct a systematic
method to obtain five-dimensional solutions to the full
system of equations governing the geometry and dynamics of the
bulk. This is done for an arbitrary bulk scalar field potential
and taking into account the presence of matter on the branes. The
method, valid in the low energy regime, is a linear
expansion of the system about the static vacuum solution.
Additionally, we develop a four-dimensional effective theory
describing the evolution of the system. At the lowest order in the
expansion, the effective theory is a bi-scalar tensor
theory of gravity. One of the main features of this theory is that
the scalar fields can be stabilized naturally without the introduction
of additional mechanisms, allowing satisfactory agreement between
the model and current observational constraints. The special case of
the Randall-Sundrum model is discussed.
\end{abstract}

\maketitle


\section{Introduction} \label{sec: Intro}

Brane-world models have become an important subject of research in
recent years. The simple and intuitive notion that the standard
model of physics is embedded in a hypersurface (brane), located
in a higher dimensional space (bulk), has significant
implications for gravity, high energy physics and cosmology. It is
strongly motivated by recent developments in string theory and its
extended version, M-theory \cite{Harova-Wiiten 1, Harova-Wiiten 2, Lukas etal}.
Additionally, it has been proposed on
a phenomenological basis \cite{Akama, Rubakov, Visser} to address many questions which cannot be
answered within the context of the conventional standard model of
physics, such as the hierarchy problem and the cosmological
constant problem \cite{Antoniadis etal, Antoniadis, RS1, RS2, Arkani etal, Forste etal}.

The basic idea is that gravity -and possibly other forces-
propagate freely in the bulk, while the Standard Model's fields
are confined to a 4-dimensional brane (see \cite{brax-bruck, Langlois, Maartens, brax-bruck-davis}
for comprehensive reviews on
brane-world models). In this type of model, constraints on the
size of extra dimensions are much weaker than in Kaluza-Klein
theories \cite{Kaluza-Klein}, although, Newton's law of gravity is
still sensitive to the presence of extra dimensions. Gravity has
been tested only to a tenth of millimeter \cite{Hoyle etal},
therefore, possible
deviations to Newton's law, below that scale, can be envisaged. In
general, one of the main modifications to general relativity
offered by brane models, is the appearance, in the effective
Einstein's equations of the system, of novel terms that depend
quadratically on the energy-momentum tensor of the matter content
of the observed Universe. When these terms can be neglected the
brane system is in the low energy regime. It is well known that
the most recent cosmological era must be considered within the low
energy regime. Despite this fact, there are still important
modifications to general relativity that must be considered, and
current observational constraints are highly relevant for the
phenomenology of brane models.

Cosmologically speaking, the evolution of the branes in the bulk
is directly related to the evolution of the observed Universe
\cite{Flanagan etal 1, Bruck etal}. So far,
most of cosmological considerations of brane-worlds have been
centered on the models proposed by Randall and Sundrum \cite{RS1,
RS2}. They considered warped geometries in which the bulk-space is
a slice of Anti de Sitter space-time. One of the main problems of
these scenarios is the stabilization of new extra degrees of
freedom appearing from the construction. More precisely, when the
four-dimensional equations of motion for the branes are
considered, a new scalar degree of freedom, the radion, emerges.
The stabilization of the radion is an important phenomenological
problem, and many different mechanisms to stabilize it have been
proposed \cite{Goldberger-Wise, Lesgourgues etal}. Clearly, brane models which only
consider the presence of gravity in the bulk, such as the
Randall-Sundrum model, are not fully satisfactory from a
theoretical point of view. There is no reason {\it a priori} for
gravity to be the only force propagating in the bulk, and other
unmeasured forces may also be present. For instance, string theory
predicts the existence of the dilaton field which acts as a
massless partner of the graviton. If such forces are present in
the bulk, they should be suppressed by some stabilization
mechanism in order to evade observation (similar to the case of the
radion field in the Randall-Sundrum model). The problem of finding
a stabilization mechanism is an active field of research.

In this paper we are going to consider a fairly general
brane-world model (BPS-branes), which has been motivated as a
supersymmetric extension of the Randall--Sundrum model
\cite{Bergshoeff etal, Binetruy etal, Davis & Brax, Csaki etal,
Youm} (see \cite{Flanagan etal 2, DeWolfe etal} for phenomenological
motivations). It consists of a five-dimensional bulk space with a
scalar field $\psi$, bounded by two branes, $\sigma_{1}$ and
$\sigma_{2}$. The main property of this system is a special
boundary condition that holds between the branes and the bulk
fields, which allows the branes to be located anywhere in the
background without obstruction. In particular, a special relation exists between the
scalar field potential $U(\psi)$, defined in the bulk, and the
brane tensions $U_{B} (\psi^{1})$ and $U_{B} (\psi^{2})$ defined
at the position of the branes. This is the
BPS condition. Due to the complexity of this set up, the main
efforts of previous works have been focused on the special case
$U_{B}(\psi) \propto e^{\alpha \psi}$, where $\alpha$ is a
parameter of the theory \cite{Br-CvdB-ACD-Rh, Kobayashi-Koyama}.
This particular potential is well motivated by supergravity in
singular spaces and string theory. For instance, when $\alpha=0$,
the Randall-Sundrum model is retrieved, meanwhile, for larger
values, $\alpha=O(1)$, one obtains the low--energy effective
action of heterotic $M$--theory, taking into account the volume of
the Calabi--Yau manifold \cite{Lukas}. Nevertheless, other classes
of potentials also have physical motivation and should be
considered. For instance, in 5-dimensional supergravity, one can
expect the linear combination $U_{B}(\psi) = V_{1} e^{\alpha_{1}
\psi} + V_{2} e^{\alpha_{2} \psi}$, where $\alpha_{1}$ and
$\alpha_{2}$ are given numbers \cite{Davis & Brax}. It must be
stated, however, that up to now the problem of assuming an
arbitrary potential $U_{B}(\psi)$ has not been considered in
detail, and therefore, the low energy regime of this type of
system is not well understood. For example, it is not known how
the gravitational fields nor the bulk scalar field behave near the
branes, and the general belief is that these issues should be
investigated via numerical techniques.

Here we study the low energy regime of brane-models with a bulk
scalar field. One of the most important aspects is that we shall
not restrict our treatment to any particular choice of the BPS
potential, allowing a clearer understanding of 5-dimensional BPS
systems, and focus our study on the problems mentioned previously.
In particular we shall develop a systematic method to obtain
5-dimensional solutions for the gravitational fields and the bulk
scalar fields. Additionally, by adopting the projective approach,
we develop an effective 4-dimensional theory, valid in the low
energy regime, and indicate that it is equivalent to the
moduli-space approximation. This effective theory is a bi-scalar
tensor theory of gravity of the form:
\begin{eqnarray}
S &=& \frac{1}{k \kappa_{5}^{2}} \int d^{4}x \sqrt{- g} \bigg[ R -
\frac{3}{4} g^{\mu \nu} \gamma_{a b}
\partial_{\mu} \psi^{a} \partial_{\nu} \psi^{b}
- \frac{3}{4} V  \bigg]  \nonumber\\
&& + S_{1}[\Psi_{1}, A_{1}^{2} g_{\mu \nu}] + S_{2}[\Psi_{2},
A_{2}^{2} g_{\mu \nu}], \label{eq1: Effective Th}
\end{eqnarray}
where $a = 1,2$ labels the branes, $\psi^{a}$ corresponds to
the projected bulk scalar field at the brane $\sigma_{a}$,
$\gamma_{a b}$ is a $\sigma$-model metric of the bi-scalar theory,
$S_{1}$ and $S_{2}$ are the actions for matter fields $\Psi_{1}$
and $\Psi_{2}$ living in the respective branes, and $A_{1}$ and
$A_{2}$ are warp factors dependent on the scalar fields. Obtaining
this effective theory is a great achievement that allows the study
of this class of models within the approach and usual techniques of
multi-scalar tensor theories \cite{Damour-Esposito, Damour}.

This paper is organized as follows: In Section \ref{Brane-Scalar},
we review brane systems with a bulk scalar field. There, we
illustrate in a more technical way some of the problems that this
paper addresses. In Section \ref{sec: Satic}, we work out and
analyze the static solutions for the vacuum configuration of
BPS-branes. This enables us to construct the low energy regime
expansion in Section \ref{sec: LER-expansion}. Then, in Section
\ref{sec: LER-eff}, we develop the four-dimensional effective
theory valid in the low energy regime. There, we will focus our
attention on the zeroth order effective theory which is a
bi-scalar tensor theory of gravity. In Section \ref{sec:
Cosmology}, we study the cosmology of the effective theory. For
instance, we consider the constraints coming from current
observations on deviations to General Relativity and analyze the
stabilization problem of BPS-systems. It will be found that, in
order to stabilize this type of system it is enough to have the
appropriate potential $U_{B}$, and therefore no exotic mechanisms
are required in general. The conclusions are summarized in Section
\ref{sec: Conclusions}.


\section{BPS-Branes with a Bulk Scalar Field}
\label{Brane-Scalar}

We start this section with the introduction of 5-dimensional
BPS configurations, deriving the 5-dimensional equations of
motion governing the dynamics of the system and deducing the
projected equations at the position of the branes.
These equations depend on the configuration of the bulk as
well as on the matter content in the branes; we analyze the
difficulties arising from this dependence to find solutions to
realistic cosmological configurations.

\subsection{Basic Configuration}

\begin{figure}[ht]
\begin{center}
\includegraphics[width=0.45\textwidth]{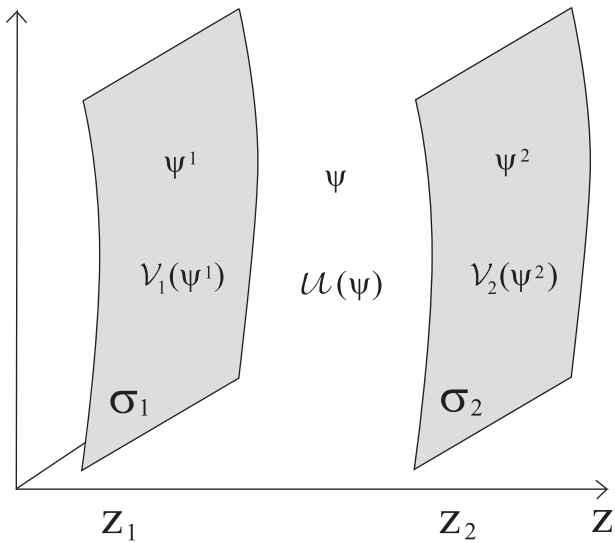}
\caption[Basic configuration]{Schematic representation of the
5-dimensional brane configuration. In the bulk there is a scalar
field $\psi$ with a bulk potential $\mathcal{U}(\psi)$.
Additionally, the bulk-space is bounded by branes, $\sigma_{1}$
and $\sigma_{2}$, with tensions given by
$\mathcal{V}_{1}(\psi^{1})$ and $\mathcal{V}_{2}(\psi^{2})$
respectively, where $\psi^{1}$ and $\psi^{2}$ are the boundary
values of $\psi$.} \label{F1}
\end{center}
\end{figure}

Let us consider a 5-dimensional manifold $M$ provided with a
coordinate system $X^{A}$, with $A=0,\ldots,4$. We shall assume
the special topology $M = \sigma \times S^{1}/ Z_{2}$, where
$\sigma$ is a fixed 4-dimensional lorentzian manifold without
boundaries and $S^{1}/Z_{2}$ is the orbifold constructed from the
1-dimensional circle with points identified through a
$Z_{2}$-symmetry. The manifold $M$ is bounded by two branes
located at the fixed points of $S^{1}/Z_{2}$. Let us denote the
brane-surfaces by $\sigma_{1}$ and $\sigma_{2}$ respectively. We
shall usually refer to the space bounded by the branes as the bulk
space. We will consider the existence of a bulk scalar field
$\psi$ living in $M$ with boundary values, $\psi^{1}$
and $\psi^{2}$, at the branes. Consider also the presence of a
bulk potential $\mathcal{U}(\psi)$ and brane tensions
$\mathcal{V}_{1}(\psi^{1})$ and $\mathcal{V}_{2}(\psi^{2})$ (which
are potentials for the boundary values $\psi^{1}$ and $\psi^{2}$).
Additionally, we will consider the existence of matter fields
$\Psi_{1}$ and $\Psi_{2}$ confined to the branes. Figure \ref{F1}
shows a schematic representation of the present configuration.

Given the present topology, it is appropriate to introduce
foliations with a coordinate system $x^{\mu}$ describing $\sigma$
(as well as the branes $\sigma_{1}$ and $\sigma_{2}$) where
$\mu=0,\ldots,3$. Additionally, we can introduce a coordinate $z$
describing the $S^{1}/Z_{2}$ orbifold and parameterizing the
foliations. With this decomposition the following form of the line
element can be used to describe $M$:
\begin{eqnarray}
ds^{2} = (N^{2} + N^{\rho} N_{\rho}) dz^{2} + 2 N_{\mu} dx^{\mu}
dz + g_{\mu \nu} dx^{\mu} dx^{\nu} \! .
\end{eqnarray}
Here $N$ and $N^{\mu}$ are the lapse and shift functions for the
extra dimensional coordinate $z$ and, therefore, they can be
defined up to a gauge choice (for example, the gaussian normal
coordinate system is such that $N^{\mu} = 0$). Additionally,
$g_{\mu \nu}$ is the pullback of the induced metric on the
4--dimensional foliations, with the $(-,+,+,+)$ signature. The
branes are located at the
fixed points of the $S^{1}/Z_{2}$ orbifold,
denoted by $z = z_{1}$ and $z =
z_{2}$, where $1$ and $2$ are the labels for the first and second
branes, $\sigma_{1}$ and $\sigma_{2}$. Without loss
of generality, we shall take $z_{1} < z_{2}$.

The total action of the system is
\begin{eqnarray} \label{eq2: Tot-Act}
S_{\mathrm{tot}} = S_{\mathrm{G}} + S_{\mathrm{bulk}} +
S_{\mathrm{BR}},
\end{eqnarray}
where $S_{\mathrm{G}}$ is the action describing the pure
gravitational part and is given by $S_{\mathrm{G}} =
S_{\mathrm{EH}} + S_{\mathrm{GH}}$, with $S_{\mathrm{EH}}$ the
Einstein--Hilbert action and $S_{\mathrm{GH}}$ the
Gibbons--Hawking boundary terms; $S_{\mathrm{bulk}}$ is the action
for the matter fields living in the bulk, which can be decomposed
into \hbox{$S_{\mathrm{bulk}} = S_{\psi} + S_{\mathrm{m}}$. Here
$S_{\psi}$} is the action for the bulk scalar field $\psi$
(including its boundary terms), while $S_{\mathrm{m}}$ is the
action for other matter fields (in the present work we will not
specify this part of the theory). Finally $S_{\mathrm{BR}}$ is the
action for the matter fields, $\Psi_{1}$ and $\Psi_{2}$, confined
to the branes.

Let us see in more detail the different contributions appearing in
equation (\ref{eq2: Tot-Act}). First, note that with the present
parameterization for $M$, we can write an infinitesimal proper
element of volume as $dV = dz d^{4}x \sqrt{-g} N$. Then, it is
possible to express $S_{\mathrm{G}}$ in the following way
\begin{eqnarray} \label{eq2: Sg}
S_{\mathrm{G}} &=& S_{\mathrm{EH}} + S_{\mathrm{GH}} \\
&=& \frac{1}{2 \kappa_{5}^{2}} \int_{S^{1}\!/Z_{2}} \!\!\!\!\!\!\!
dz \int_{\sigma} \!\! d^{\, 4} x \sqrt{-g} \, N \big( R - [ K_{\mu
\nu} K^{\mu \nu} - K^{2} ] \big), \nonumber
\end{eqnarray}
where $R$ is the four-dimensional Ricci scalar constructed from
$g_{\mu \nu}$, and $\kappa_{5}^{2} = 8 \pi G_{5}$, with $G_{5}$
the five-dimensional Newton's constant. Additionally,
\hbox{$K_{\mu \nu} = \left[ g_{\mu \nu}' - \nabla_{\mu} N_{\nu} -
\nabla_{\nu} N_{\mu} \right] / 2N$} is the extrinsic curvature of
the foliations and $K$ its trace (the prime denotes
derivatives in terms of $z$, that is $' = \partial_{z}$, and
covariant derivatives, $\nabla_{\mu}$, are constructed from the
induced metric $g_{\mu \nu}$ in the standard way). On the other
hand the action for the bulk scalar field, $S_{\psi}$, can be
written in the form
\begin{eqnarray}
S_{\psi} &=& - \frac{3}{8 \kappa_{5}^{2}} \int dz \, d^{4} x
\sqrt{-g} \, N \bigg[ \bigg(\frac{\psi'}{N} \bigg)^{2}
+ (\partial \psi)^{2}  \nonumber \\
&& - 2 \frac{N^{\mu}}{N^{2}} \,
\partial_{\mu} \psi \, \partial_{z} \psi  + \mathcal{U}(\psi) \bigg] +
S_{\psi}^{\,1} + S_{\psi}^{\,2} \!,
\end{eqnarray}
where $S_{\psi}^{\,1}$ and  $S_{\psi}^{\,2}$ are boundary terms
given by
\begin{eqnarray} \label{eq2: brane-tensions}
S_{\psi}^{\,1} &=& - \frac{3}{2 \kappa_{5}^{2}} \int_{\sigma_{1}}
\!\!\! d^{4} x \sqrt{-g} \, \mathcal{V}_{1}(\psi^{1}), \\
S_{\psi}^{\,2} &=& + \frac{3}{2 \kappa_{5}^{2}} \int_{\sigma_{2}}
\!\!\! d^{4} x \sqrt{-g} \, \mathcal{V}_{2}(\psi^{2}),
\end{eqnarray}
at the respective positions, $z_{1}$ and $z_{2}$. In the above
expressions $\mathcal{U}(\psi)$ is the bulk scalar field
potential, while $\mathcal{V}_{1}(\psi^{1})$ and
$\mathcal{V}_{2}(\psi^{2})$ are boundary potentials defined at the
positions of the branes (also referred as the brane tensions). In
the present case (BPS-configurations), we shall consider the
following general form for the potentials:
\begin{eqnarray}
\mathcal{U} &=& U + u, \\\mathcal{V}_{1} &=& U_{B} + v_{1}, \\
\mathcal{V}_{2} &=& U_{B} + v_{2},
\end{eqnarray}
where $U$ and $U_{B}$ are the bulk and brane superpotentials, and
the potentials $u$, $v_{1}$ and $v_{2}$ are such that
\mbox{$|u| \ll |U|$} and $|v_{1}|, |v_{2}| \ll |U_{B}|$. In this
way, the system is dominated by the superpotentials $U$ and
$U_{B}$.  The most important characteristic of this class of
system is the relation between $U$ and $U_{B}$ (the BPS-relation),
given by:
\begin{eqnarray} \label{eq2: BPS-cond}
U= \left( \partial_{\psi} U_{B}  \right)^{2} - U_{B}^{2}.
\end{eqnarray}
As we shall see later in more detail, the BPS property mentioned
earlier (the fact that the branes can be located anywhere in the
background, without obstruction) comes from the relation above.
This specific configuration, when the potentials $u, v_{1}, v_{2}
= 0$ and no fields other than the bulk scalar field are present,
is referred as the BPS configuration. It emerges, for instance, in
models like $N=2$ supergravity with vector multiplets in the bulk
\cite{Davis & Brax}. More generally, it can be regarded as a
supersymmetric extension of the Randall-Sundrum model. In
particular when $U_{B}$ is the constant potential, the
Randall--Sundrum model is recovered with a bulk cosmological
constant $\Lambda_{5} = (3/8) U =  - (3/8) \, U_{B}^{2}$, and the
usual fine tuning condition is given by equation (\ref{eq2:
BPS-cond}). The presence of the potentials $u$, $v_{1}$ and
$v_{2}$ are generally expected from supersymmetry breaking
effects. However, their specific form is not known and, up to now,
they must be arbitrarily postulated \cite{brax-chatillon}.

Finally, for the matter fields confined to the branes, we shall
consider the standard action:
\begin{eqnarray}
S_{\mathrm{BR}} = S_{1}[\Psi_{1},g_{\mu \nu}(z_{1})] +
S_{2}[\Psi_{2},g_{\mu \nu}(z_{2})],
\end{eqnarray}
where $\Psi_{1}$ and $\Psi_{2}$ denote the respective matter
fields, and $g_{\mu \nu} (z_{a})$ is the induced metric at
position $z_{a}$. In this article we shall assume that the matter
fields are minimally coupled to the bulk scalar field. That is to
say, we will assume that $S_{\mathrm{BR}}$ does not depend
explicitly on $\psi$. Nevertheless, it is important to point out
that, in general, one should expect some nontrivial coupling
between the bulk scalar field $\psi$ and the matter fields
$\Psi_{1}$ and $\Psi_{2}$, that could lead to the variation of
constants in the brane-world scenario \cite{Palma etal}. Despite
these comments, we show
that, in a minimally coupled system, an effective coupling is
generated between the matter fields and the bulk scalar field.

\subsection{Equations of Motion}

In this subsection we derive the equations of motion governing the
dynamics of the fields living in the bulk. These equations are
obtained by varying the action $S_{\mathrm{tot}}$ with respect to
the gravitational and bulk scalar fields, without taking
into account the boundary terms. These will be
considered in the next subsection, in order to deduce a set of
boundary conditions for the bulk fields.

The 5-dimensional Einstein's equations of the system can be
derived through the variation of the total action
$S_{\mathrm{tot}}$ with respect to $N$, $N_{\mu}$ and $g_{\mu
\nu}$.  Respectively, these equations are found to be:
\begin{eqnarray}
R + \left[ K_{\mu \nu} K^{\mu \nu} - K^{2} \right] = 2
\kappa_{5}^{2} \, \tilde S, \label{eq2: Grav-eq1}
\\
\nabla_{\mu} \left[ K^{\mu}_{\nu} - K \delta^{\mu}_{\nu} \right] =
- \kappa_{5}^{2} \, \tilde J_{\nu}, \label{eq2: Grav-eq2}
\\
G_{\mu \nu} + E_{\mu \nu} = \frac{1}{2} g_{\mu \nu} \left[ K_{\rho
\sigma} K^{\rho \sigma} - K^{2} \right]  + K K_{\mu \nu}
\nonumber\\ - K^{\sigma}_{\mu} K_{\sigma \nu} + \frac{2}{3}
\kappa_{5}^{2} \bigg[ \tilde T_{\mu \nu} - \frac{1}{4} g_{\mu \nu}
(\tilde T_{\sigma}^{\sigma} + 3 \tilde S) \bigg]. \label{eq2:
Grav-eq3}
\end{eqnarray}
The above set of equations deserves a detailed description.
$G_{\mu \nu}$, in (\ref{eq2: Grav-eq3}), is the Einstein tensor
constructed from the induced metric $g_{\mu \nu}$. The quantities
$\tilde T_{\mu \nu}$, $\tilde J_{\mu}$, and $\tilde S$ are the
4-dimensional stress energy-momentum tensor, vector and scalar
respectively, characterising the matter content in the bulk space.
They are defined as
\begin{eqnarray}
\tilde S &=& -
\frac{1}{\sqrt{-g}} \frac{\delta S_{\mathrm{bulk}}}{\delta N},
\nonumber\\
\tilde J_{\mu} &=& - \frac{1}{\sqrt{-g}} \frac{\delta
S_{\mathrm{bulk}}}{\delta N^{\mu}}, \nonumber\\
\tilde T_{\mu \nu} &=& - \frac{2}{N \sqrt{-g}} \frac{\delta
S_{\mathrm{bulk}}}{\delta g^{\mu \nu}}, \label{eq2: tilde-S-J-T}
\end{eqnarray}
where $S_{\mathrm{bulk}} = S_{\psi} + S_{\mathrm{m}}$ is the
action for the matter fields living in the bulk. The above
quantities are decompositions of the usual 5-dimensional stress
energy-momentum tensor, $T_{A B}$. To be more specific, the
definitions in (\ref{eq2: tilde-S-J-T}) satisfy: \mbox{$\tilde
T_{\mu \nu} = X_{, \, \mu}^{A} \, X_{, \, \nu}^{B} \, T_{A B}$},
$\tilde J_{\mu} = - X_{, \, \mu}^{A} \, n^{B} \, T_{A B}$, and
$\tilde S = - n^{A} n^{B} \, T_{A B}$, where $X_{, \, \mu}^{A}$ is
the pullback between the coordinate system $x^{\mu}$ of the
foliations and the 5-dimensional coordinate system $X^{A}$ of $M$,
and where $n^{A}$ is the unit-vector field normal to the
foliations. In particular, the contributions coming from the
scalar field action, $S_{\psi}$, are given by:
\begin{eqnarray}
\tilde S^{(\psi)} &=& - \frac{3}{8 \kappa_{5}^{2}} \bigg[
\frac{1}{N^{2}} (\psi')^{2} - 2 \frac{N^{\mu}}{N^{2}} \,
\partial_{\mu} \psi \, \partial_{z} \psi - (\partial \psi)^{2}
- \mathcal{U} \bigg], \nonumber\\
  \tilde J^{(\psi)}_{\mu} &=& - \frac{3}{4 \kappa_{5}^{2}}
\frac{1}{N} \,
\partial_{z} \psi
\partial_{\mu} \psi,
\nonumber\\
\tilde T^{(\psi)}_{\mu \nu} &=& - \frac{3}{4 \kappa_{5}^{2}}
\bigg[ \frac{1}{2} g_{\mu \nu} \bigg(  \frac{1}{N^{2}} (\psi')^{2}
- 2 \frac{N^{\mu}}{N^{2}} \,
\partial_{\mu} \psi \, \partial_{z} \psi
\nonumber\\
&& + (\partial \psi)^{2}  + \mathcal{U} \bigg)
- \partial_{\mu} \psi \partial_{\nu} \psi \bigg].
\end{eqnarray}
Also in (\ref{eq2: Grav-eq3}) is $E_{\mu \nu}$, the projection of the
5-dimensional Weyl tensor, $C^{C}_{\,\,\,\, A D B}$, on the foliations
given by $E_{\mu \nu} = X_{, \, \mu}^{A} X_{, \, \nu}^{B} \, n_{C} \, n^{D} \,
C^{C}_{\,\,\, A D B}$. In the present set up, this can be written as:
\begin{eqnarray}
E_{\mu \nu} &=& K_{\mu}^{\alpha} K_{\alpha \nu}  - \frac{1}{N}
\left( \nabla_{\mu} \nabla_{\nu} N  + \pounds_{N n} K_{\mu \nu}
\right) \nonumber\\ && - \frac{1}{3} \kappa_{5}^{2} \left[ \tilde
T_{\mu \nu} - \frac{1}{2} g_{\mu \nu} (\tilde T_{\sigma}^{\sigma}
+ \tilde S) \right],
\end{eqnarray}
where $\pounds_{N n}$ is the Lie derivative along the vector field
$N n^{A}$. At this point it is convenient to adopt the gauge
choice $N^{\mu} = 0$, which corresponds to the choice of
gaussian normal coordinates. With this gauge, one has $\pounds_{N n} =
\partial_{z}$, and the treatment of the entire system is greatly
simplified. Next, we can obtain the equation of motion for the
bulk scalar field $\psi$. Varying the action with respect to $\psi$ gives:
\begin{eqnarray} \label{eq2: scalar}
g^{\mu \nu} \nabla_{\mu} \left( N
\partial_{\nu} \psi \right) + \left( \psi' / N \right)' +
K \psi'  = \frac{N}{2} \frac{\partial
\mathcal{U}}{\partial \psi}.
\end{eqnarray}
(Recall that we are now using $N^{\mu} = 0$).

It is possible to see that equations (\ref{eq2: Grav-eq1}) and
(\ref{eq2: Grav-eq3}) are the dynamical equations for the
gravitational sector (notice that second order derivatives in
terms of $z$ are only present in the projected Weyl tensor $E_{\mu
\nu}$). Meanwhile, equation (\ref{eq2: Grav-eq2}) constitutes a
restriction for the extrinsic curvature $K_{\mu \nu}$ in terms of
the matter content of the bulk. As we shall soon see in more
detail, the conservation of energy-momentum on the branes is a
direct consequence of equation (\ref{eq2: Grav-eq2}).

\subsection{Boundary Conditions}

We can now turn to the boundaries of the system. The former
variations leading to equations (\ref{eq2:
Grav-eq1})-(\ref{eq2: Grav-eq3}) and (\ref{eq2: scalar}) also
give rise to boundary conditions that must be respected by the
bulk fields at positions $z_{1}$ and $z_{2}$. They emerge when we
take into account the boundary terms present in the action
(including the matter fields in the branes). For the first brane,
$\sigma_{1}$, these conditions are found to be:
\begin{eqnarray}
\{ K_{\mu \nu} - g_{\mu \nu} K \}|_{z_{1}} &=& \frac{3}{2} \left[
U_{B} + v_{1} \right] \, g_{\mu \nu}
 - \kappa_{5}^{2} T_{\mu \nu}^{1}, \quad \label{eq2: Israel-matching-s1}\\
  \frac{1}{2} \{ \psi'/N \}|_{z_{1}} &=&
\partial_{\psi} \left[ U_{B} + v_{1} \right], \label{eq2: BPS-matching-s1}
\end{eqnarray}
while for the second brane, $\sigma_{2}$, these conditions are:
\begin{eqnarray}
\{ K_{\mu \nu} - g_{\mu \nu} K \}|_{z_{2}} &=& - \frac{3}{2}
\left[ U_{B} + v_{2} \right] \, g_{\mu \nu}
 - \kappa_{5}^{2} T_{\mu \nu}^{2}, \quad \label{eq2: Israel-matching-s2}\\
  \frac{1}{2} \{ \psi'/N \}|_{z_{2}} &=& -
\partial_{\psi} \left[ U_{B} + v_{2} \right]. \label{eq2: BPS-matching-s2}
\end{eqnarray}
Equations (\ref{eq2: Israel-matching-s1}) and (\ref{eq2:
Israel-matching-s2}) are the Israel matching conditions
and equations (\ref{eq2: BPS-matching-s1}) and (\ref{eq2:
BPS-matching-s2}) are the BPS matching conditions for the bulk
scalar field. In the previous expressions the brackets denote the
difference between the evaluation of quantities at both sides of
the branes, i.e $\{ f \}|_{z} = \lim_{\epsilon \rightarrow 0} [f(z
+ \epsilon) - f(z - \epsilon)]$. Additionally, $T^{a}_{\mu \nu}$
denotes the energy--momentum tensors of the matter content in the
brane $\sigma_{a}$, and is defined by the standard expression:
\begin{eqnarray}
T^{a}_{\mu \nu} = - \frac{2}{\sqrt{-g}}\frac{\delta S_{a}}{\delta
g^{\mu \nu}}
\bigg|_{z_{a}}.
\end{eqnarray}
Since we are considering an orbifold with a $Z_{2}$-symmetry, and
the branes are positioned at the fixed points, the conditions
(\ref{eq2: Israel-matching-s1})-(\ref{eq2: BPS-matching-s2}) can
be rewritten as
\begin{eqnarray}
 K_{\mu \nu} - g_{\mu \nu} K &=&  \frac{3}{4} \left[ U_{B}
 + v_{1} \right] \, g_{\mu \nu}
  - \frac{\kappa_{5}^{2}}{2} T_{\mu \nu}^{1}, \label{eq2: Israel-matching2-s1} \\
\psi' &=& N \, \partial_{\psi} \left[ U_{B} + v_{1} \right],
\label{eq2: BPS-matching2-s1}
\end{eqnarray}
for the first brane, $\sigma_{1}$, and
\begin{eqnarray}
 K_{\mu \nu} - g_{\mu \nu} K &=&  \frac{3}{4} \left[ U_{B}
 + v_{2} \right] \, g_{\mu \nu}
  + \frac{\kappa_{5}^{2}}{2} T_{\mu \nu}^{2}, \label{eq2: Israel-matching2-s2} \\
\psi' &=& N \, \partial_{\psi} \left[ U_{B} + v_{2} \right],
\label{eq2: BPS-matching2-s2}
\end{eqnarray}
for the second brane, $\sigma_{2}$. These conditions are of utmost
importance. They relate the geometry of the bulk with the matter
content in the branes. In the next subsection we shall examine in
more detail how these conditions are related to the 4-dimensional
effective equations in the branes and analyze some of the
open questions.

\subsection{Short Analysis and Open Questions}

Equations (\ref{eq2: Grav-eq1})-(\ref{eq2: Grav-eq3}) and
(\ref{eq2: scalar}) can be evaluated at the position of the
branes, with the help of the boundary conditions (\ref{eq2:
Israel-matching2-s1})-(\ref{eq2: BPS-matching2-s2}), to obtain a
set of 4-dimensional effective equations. This is the projective
approach. For instance, evaluating equations (\ref{eq2: Grav-eq1})
and (\ref{eq2: Grav-eq3}) at the position of the first brane, we
find:
\begin{eqnarray}
\lefteqn{R = - \frac{\kappa_{5}^{2}}{4} \mathcal{V}_{1} T
- g^{\mu \nu} \Pi_{\mu \nu}  + \frac{3}{4} W } \nonumber\\
&& {} + \frac{3}{4} (\partial \psi)^{2}
- g^{\mu \nu} \Theta^{(\mathrm{m})}_{\mu \nu} ,   \label{eq2: Proj-R1}  \\
\lefteqn{G_{\mu \nu} = \frac{\kappa_{5}^{2}}{4} \mathcal{V}_{1}
T_{\mu
\nu} + \Pi_{\mu \nu} + \Theta^{(\mathrm{m})}_{\mu \nu} - E_{\mu \nu}} \nonumber \\
&& {} + \frac{1}{2} \left[
\partial_{\mu} \psi \partial_{\nu} \psi - \frac{5}{8} g_{\mu \nu}
(\partial \psi)^{2}
 \right] - \frac{3}{16} g_{\mu \nu} W, \label{eq2: Proj-Ein1}
\end{eqnarray}
where $\Theta_{\mu \nu}$, $\Pi_{\mu \nu}$ and $W$ have been
defined as:
\begin{eqnarray}
\Theta_{\mu \nu}^{\mathrm{(m)}} &=& \frac{2}{3} \kappa_{5}^{2}
\bigg[ \tilde T^{\mathrm{(m)}}_{\mu \nu}
 - \frac{1}{4}
g_{\mu \nu} \left(\tilde T_{\,\,\,\,\,\,\,\, \sigma}^{\mathrm{(m)}
\,
\sigma} + 3 \tilde S^{\mathrm{(m)}} \right) \bigg], \qquad \\
\Pi_{\mu \nu} &=& \frac{\kappa_{5}^{4}}{4} \bigg[ \frac{1}{3} T \,
T_{\mu \nu} - T_{\mu \sigma} T^{\sigma}_{\nu} \nonumber\\ && +
\frac{1}{2} g_{\mu \nu} T_{\sigma \rho} T^{\sigma \rho} -
\frac{1}{6} g_{\mu \nu}
T^{2} \bigg], \\
W &=& \mathcal{U} - \bigg( \frac{\partial
\mathcal{V}_{1}}{\partial \psi} \bigg)^{2}
 + \mathcal{V}_{1}^{2}.
\end{eqnarray}
Notice that we have dropped the index labelling the first brane in
the energy momentum tensor (that is, we have taken $T_{\mu \nu} =
T^{1}_{\mu \nu}$ and $T = g^{\mu \nu} T^{1}_{\mu \nu}$).
Similarly, the projection of equation (\ref{eq2: scalar}) on the
first brane leads us to the 4-dimensional effective bulk scalar
field equation. This is found to be:
\begin{eqnarray} \label{eq2: Proj-scalar}
\Box \psi = - \frac{1}{N}
\partial^{\mu} N \partial_{\mu} \psi - \frac{\kappa_{5}^{2}}{6} \frac{\partial
\mathcal{V}_{1}}{\partial \psi} \, T - \Delta \psi + \frac{1}{2}
\frac{\partial u}{\partial \psi} \nonumber\\ + \left[
\frac{\partial U_{B}}{\partial \psi} \, v_{1} + \frac{\partial
v_{1}}{\partial \psi} \bigg( \mathcal{V}_{1} - \frac{\partial^{2}
U_{B}}{\partial \psi^{2}} \bigg) \right] ,
\end{eqnarray}
where we have defined the loss parameter $\Delta \psi$ of the
system as:
\begin{eqnarray}
\Delta \psi = \frac{1}{N} [\psi' / N -
\partial_{\psi} U_{B}]\,'.
\end{eqnarray}
Additionally, from the evaluation of equation (\ref{eq2:
Grav-eq2}) at position $z_{1}$, one finds the energy-momentum
conservation relation for the matter fields of the first brane:
\begin{eqnarray} \label{eq2: cons-T}
\nabla_{\mu} T^{\mu}_{\nu} =  2 \tilde J^{\mathrm{(m)}}_{\nu} .
\end{eqnarray}
Note that if $u = v_{1} = 0$ then $W = 0$. The precise form of
$\tilde T_{\mu \nu}^{\mathrm{(m)}}$, $\tilde
J^{\mathrm{(m)}}_{\nu}$ and $\tilde S^{\mathrm{(m)}}$, in
equations (\ref{eq2: Proj-R1}), (\ref{eq2: Proj-Ein1}) and
(\ref{eq2: cons-T}), depend on the bulk matter fields considered
in $S_{\mathrm{m}}$. From now on we will not consider any
contribution from $S_{\mathrm{m}}$ and therefore we shall take
$\tilde T_{\mu \nu}^{\mathrm{(m)}} = \tilde J^{\mathrm{(m)}}_{\nu}
= \tilde S^{\mathrm{(m)}} = 0$ (that is, the only fields living in
the bulk are the gravitational fields $g_{\mu \nu}$ and $N$, and
the bulk scalar field $\psi$).

Equations (\ref{eq2: Proj-R1}), (\ref{eq2: Proj-Ein1}) and
(\ref{eq2: Proj-scalar}) are the projected equations describing
the theory in the first brane (an analog set of equations can be
obtained for the second brane). To solve them, it is important to
know the precise form of the projected Weyl tensor $E_{\mu \nu}$
and the loss parameter $\Delta \psi$. In order to find a complete
solution of the entire system it is necessary to find a solution
for these quantities. They contain the necessary information from the bulk, or
equivalently, from a brane frame point of view; they propagate the
information from one brane to the other. A consistent method to
compute $E_{\mu \nu}$ and $\Delta \psi$ is an open question.

Observe that the 4-dimensional effective Newton's constant,
$G_{4}$, can be identified as $8 \pi G_{4} = \kappa_{5}^{2}
\mathcal{V}_{1}/4$ and, therefore, it is a function of the
projected bulk scalar field $\psi$. Additionally, it can be seen
that there are two scalar degrees of freedom. Naturally, one them
is the projected bulk scalar field $\psi$. The second one
corresponds to the projected value of lapse function $N$, which is
normally referred as the radion field. In the absence of
supersymmetry breaking terms $u$, $v_{1}$ and $v_{2}$, these
scalar degrees of freedom are massless and are the
moduli fields. The stabilization of moduli fields is one of the
most important phenomenological issues in brane models to ensure
agreement with observations.

The term $\Pi_{\mu \nu}$, in equations (\ref{eq2: Proj-Ein1}) and
(\ref{eq2: Proj-scalar}), contains quadratic contributions from
the energy momentum tensor $T_{\mu \nu}$. Therefore, $\Pi_{\mu
\nu}$ is relevant at the high energy regime and it will turn out
to be important for the early universe physics. At more recent
times, in order to agree with observations, this term must be
negligible, which is only possible if $\kappa_{5}^{2} \, | U_{B} |
\gg |T| $. This is the low energy regime. Neglecting $\Pi_{\mu
\nu}$ makes the previous equations of motion take the form:
\begin{eqnarray} \label{eq2: Proj-low}
R &=& - \frac{\kappa_{5}^{2}}{4} U_{B} T
+ \frac{3}{4} (\partial \psi)^{2}, \\
G_{\mu \nu} &=& \frac{\kappa_{5}^{2}}{4} U_{B} \, T_{\mu \nu} -
E_{\mu \nu} \nonumber\\ && + \frac{1}{2} \bigg[
\partial_{\mu} \psi \partial_{\nu} \psi - \frac{5}{8} g_{\mu \nu}
(\partial \psi)^{2}
 \bigg], \\
\Box \psi  &=& - \frac{1}{N}
\partial^{\mu} N \partial_{\mu} \psi - \frac{\kappa_{5}^{2}}{6} \frac{\partial
U_{B}}{\partial \psi} \, T  - \Delta \psi, \label{eq2:
Proj-low-sca}
\end{eqnarray}
where we have, for simplicity, also dropped the terms containing
$u$ and $v_{1}$. Recall the additional
equation expressing energy-momentum conservation on the first brane,
which now reads:
\begin{eqnarray}
\nabla_{\mu} T^{\mu}_{\nu} = 0. \label{eq2: conserv}
\end{eqnarray}
This version for the equations of motion are much simpler than the
full system, however, the difficulties mentioned above are still
present and many questions remain to be answered. For example, the
usual approach is to neglect the effects of the loss parameter
$\Delta \psi$ and take it to be zero. If this is the case, observe
that the projected scalar field $\psi$, in equation (\ref{eq2:
Proj-low-sca}), is driven by the derivative $\partial_{\psi}
U_{B}$ times the trace of the energy-momentum tensor. One could
therefore expect the scalar field to be stabilized provided the
appropriate form of $U_{B}$. This possibility, however, will
depend on the dynamics of the radion field $N$, which, at the same
time depends on the dynamics of the inter-brane system. The
stabilization of the scalar field and the radion are usually
solved by the introduction of new potential terms for $\psi$ at
the branes and the bulk. In the present case that could be the
role of $u$, $v_{1}$ and $v_{2}$.

In the rest of the paper we shall focus our efforts to solve these
problems in the low energy regime. For instance, we will
see that it is possible to introduce a procedure to compute
$E_{\mu \nu}$ and $\Delta \psi$ with any desire precision.
Additionally, we show that it is possible to stabilize the scalar
degrees of freedom solely with the introduction of the appropriate
potential $U_{B}$ (that is, without the need of supersymmetry
breaking terms $u$, $v_{1}$ and $v_{2}$).


\section{Static Vacuum Configuration} \label{sec: Satic}

Before studying realistic solutions in detail, namely when matter
fields are present in the branes, let us analyze the static vacuum
solutions to this system. This will be highly relevant for the
development of the low energy expansion and the effective theory,
to be worked out in the next sections.

\subsection{Static Vacuum Solution}

Let us consider the presence of the bulk scalar field
$\psi$ and gravitational fields $N$ and $g_{\mu \nu}$. In this
case, we are mainly concerned with static vacuum solutions
to the system of equations (\ref{eq2: Grav-eq1})-(\ref{eq2:
Grav-eq3}) and (\ref{eq2: scalar}). Therefore, we shall assume
that $\psi$, $N$ and $g_{\mu \nu}$ are such that:
\begin{eqnarray}
\partial_{\mu} \psi = 0, \,\, \quad
\partial_{\mu} N = 0, \,\, \quad \mathrm{and} \,\, \quad
G_{\mu \nu} = 0,
\end{eqnarray}
where $G_{\mu \nu}$ is the Einstein's tensor constructed from
$g_{\mu \nu}$. To solve the system, it is sensible to consider the
following form for the induced metric:
\begin{eqnarray}
g_{\mu \nu} = \omega^{2}(z) \, \tilde g_{\mu \nu}(x),
\end{eqnarray}
where $\tilde g_{\mu \nu}(x)$ is a metric that depends only on the
space-time coordinate $x$, and which necessarily satisfies the 4-dimensional
vacuum Einstein's equations $\tilde G_{\mu \nu} = 0$. In this way,
all the dependence of the induced metric on the extra-dimensional
coordinate $z$ is contained in the warp factor $\omega(z)$. With
this form of the metric, the set of matching conditions (\ref{eq2:
Israel-matching2-s1})-(\ref{eq2: BPS-matching2-s2}) can be
rewritten in the following simple way:
\begin{eqnarray}
\frac{\omega'}{\omega} &=& - \frac{1}{4} N U_{B}, \label{eq3: match-omega} \\
\psi'  &=& N \frac{ \partial U_{B}}{\partial \psi}. \label{eq3:
match-scalar}
\end{eqnarray}
More importantly, it is possible to show that these relations
solve the entire system of equations (\ref{eq2:
Grav-eq1})-(\ref{eq2: Grav-eq3}) and (\ref{eq2: scalar}). This
important fact constitutes one of the main properties of
BPS-systems. It means that the branes can be arbitrarily located
anywhere in the background, without obstruction. It should be
clear that when matter is allowed to exist in the branes, the
boundary conditions (\ref{eq3: match-omega}) and (\ref{eq3:
match-scalar}) will not continue being solutions to the system of
equations (\ref{eq2: Grav-eq1})-(\ref{eq2: Grav-eq3}), and the
static configurations will not be possible in general. The
introduction of matter in the branes will drive the branes to a
cosmological evolution.

In the static vacuum solution expressed through equations
(\ref{eq3: match-omega}) and (\ref{eq3: match-scalar}), the
dependence of the lapse function $N$, in terms of the extra
dimensional coordinate $z$, is completely arbitrary, though it must
be restricted to be positive in the entire bulk, and its
precise form will correspond to a gauge choice. Let us assume
that $\psi(z)$ satisfies equations (\ref{eq3: match-omega}) and
(\ref{eq3: match-scalar}), and that it has boundary values $\psi^{1}$
and $\psi^{2}$ defined by:
\begin{eqnarray}
\psi^{1} = \psi(z_{1}) \qquad \mathrm{and} \qquad \psi^{2} =
\psi(z_{2}).
\end{eqnarray}
Since we are interested in the static vacuum solution, $\psi^{1}$
and $\psi^{2}$ are just constants. The precise form of $\psi(z)$,
as a function of $z$, depends on the form of $U_{B}(\psi)$ and
the gauge choice for $N$. However, it is not difficult to see that
$\psi^{1}$ and $\psi^{2}$ are the only degrees of freedom
necessary to specify the BPS state of the system. That is, given a
gauge choice for $N$, in general we have:
\begin{eqnarray}
\psi = \psi(z, \psi^{1}, \psi^{2}), \quad \mathrm{and} \quad N =
N(z, \psi^{1}, \psi^{2}).
\end{eqnarray}

Observe, additionally, that $\omega(z)$ can be expressed in terms
of $\psi(z)$ in a gauge independent way. Using both relations,
(\ref{eq3: match-omega}) and (\ref{eq3: match-scalar}), we find:
\begin{eqnarray}
\omega(z) &=& \exp \left[ -\frac{1}{4} \int_{\psi^{1}}^{\psi(z)}
\!\! \alpha^{-1}(\psi) \, d \psi \right], \label{eq3: omega} \\
\alpha(\psi) &=& \frac{1}{U_{B}} \frac{\partial U_{B}}{\partial
\psi}.
\end{eqnarray}
In the last equations we have normalized the solution $\omega(z)$ in such a
way that the induced metric to the first brane is $\tilde g_{\mu
\nu}$, though this choice is not strictly necessary. The induced metric
on the second brane is, therefore, conformally
related to the first brane, with a warp factor $\omega(z_{2})$.

Let us emphasize the fact that, given a
solution $\psi (z)$ and a gauge choice $N$, satisfying equations
(\ref{eq3: match-omega}) and (\ref{eq3: match-scalar}), the entire
system can be specified by providing the degrees of freedom
$\psi^{1}$, $\psi^{2}$ and $\tilde g_{\mu \nu}$. In the next
section we shall see that when matter fields are considered, the
vacuum solution is perturbed and $\psi^{1}$, $\psi^{2}$ and
$\tilde g_{\mu \nu}$ must be promoted to satisfy non-vacuum
equations of motion. The resulting theory will be a bi-scalar
tensor theory of gravity, with the two scalars given by $\psi^{1}$
and $\psi^{2}$.

\subsection{Bulk Geometry}

\begin{figure}[ht]
\begin{center}
\includegraphics[width=0.4\textwidth]{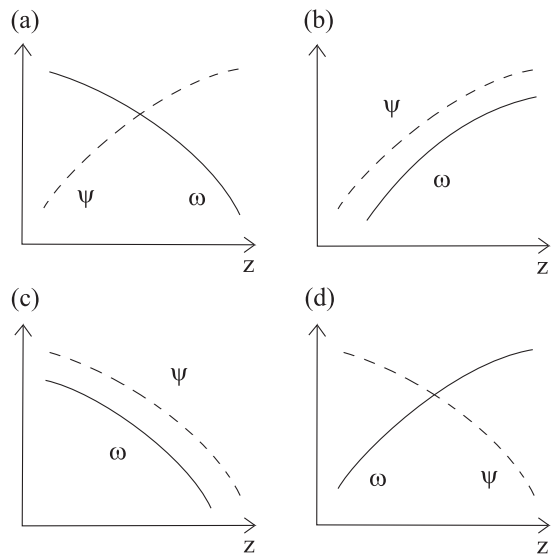}
\caption[Behaviour of the fields]{The figure shows the four
different possibilities for the behaviour of $\psi$ and $\omega$ as
a function of $z$, according to the signs of $\partial_{\psi}
U_{B}$ and $U_{B}$. The parameters of the figures are as follows:
(a) $\partial_{\psi} U_{B} > 0$ and $U_{B} > 0$ (or $\alpha > 0$).
(b) $\partial_{\psi} U_{B} > 0$ and $U_{B} < 0$ (or $\alpha < 0$).
(c) $\partial_{\psi} U_{B} < 0$ and $U_{B} > 0$ (or $\alpha < 0$).
(d) $\partial_{\psi} U_{B} < 0$ and $U_{B} < 0$ (or $\alpha >
0$).} \label{F2}
\end{center}
\end{figure}

It is instructive to know the behaviour of the scalar field $\psi$ and
warp factor $\omega$ in the bulk. For instance, a relevant
question is what the conditions are for a singularity to be
present in the bulk? Let us start by noting that equations
(\ref{eq3: match-omega}) and (\ref{eq3: match-scalar}) give us the
way in which the scalar field $\psi$ and the warp factor
$\omega$ behave as functions of the proper distance, $d \zeta = N
dz$, in the bulk space. If $U_{B}
> 0$, then the warp factor will decrease in the $z$
direction, while if $U_{B} < 0$ then the warp factor will be
increasing. Similarly, if $\partial_{\psi} U_{B} > 0$, then the
scalar field will increase in the $z$ direction, and if
$\partial_{\psi} U_{B} < 0$, then it will decrease.
Moreover, if $U_{B}(\psi^{1})>0$ then the first brane has a positive
tension while if  $U_{B}(\psi^{1})<0$ then it has a negative tension.
The situation for the second brane is similar:
if $U_{B}(\psi^{2})>0$ then it has a negative
tension while if  $U_{B}(\psi^{2})<0$ then it has a positive tension.
In this way, it is possible to see that in general the two branes will not
necessarily have opposite tensions. Figure \ref{F2} sketches the
different possible behaviours for $\psi$ and $\omega$ as functions
of $z$.

From equation (\ref{eq3: match-scalar}) we see that
the infinitesimal proper distance, $d \zeta$, in the
extra-dimensional direction can be written as:
\begin{eqnarray}
d \zeta = \left( \frac{\partial U_{B}}{\partial \psi}
\right)^{\!\! -1} \!\!\! d \psi. \label{eq3: para-z-psi}
\end{eqnarray}
Observe from this last relation that extremum points of the
potential $U_{B}$, given by the condition $\partial_{\psi} U_{B} =
0$, can be only reached at an infinite distance away in the bulk.
This is easily seen from the fact that $d \zeta \rightarrow +
\infty$ as $\partial_{\psi} U_{B} \rightarrow 0$. (The asymptotic
behaviours at the extremum points are, $\omega \rightarrow 0$ if
$U_{B}
> 0$ and $\omega \rightarrow + \infty$ if $U_{B} < 0$). Thus
$\partial_{\psi} U_{B}$ cannot change sign in the bulk space
and, therefore, $\psi(z)$ is a monotonic function of $z$. This is
not the case, however, for the warp factor $\omega$. This can
increase or decrease according to the sign of $U_{B}(\psi)$ for a
given value of $\psi$. The fact that $\psi(z)$ is monotonic in the
bulk space allows the possibility of parameterizing the bulk with
$\psi$ instead of $z$. This is an important result that will be
heavily exploited in the rest of this paper.

Finally, let us examine the possibility of having singularities at
a finite position in the bulk. From equation (\ref{eq3: omega}) we
can see that singularities will appear in the bulk whenever
$\omega(z)=0$ or $\omega(z) = + \infty$. In general, these
singularities will be located at points, $z_{\infty}$, where the
following integral diverges:
\begin{eqnarray} \label{eq3: integr-div}
\int_{\psi^{1}}^{\psi(z)} \!\!\!\!\! \alpha^{-1}(\psi) \, d \psi .
\end{eqnarray}
Let us designate $\psi_{\infty} = \psi(z_{\infty})$ to the value
of the scalar field at which (\ref{eq3: integr-div}) becomes
divergent. Then, from equation (\ref{eq3: para-z-psi}), it is
possible to see that for a singularity to be located at a finite
proper-distance in the bulk, the following additional condition
needs to be satisfied:
\begin{eqnarray} \label{eq3: cond-sing}
\Bigg| \int_{\psi^{1}}^{\psi_{\infty}} \left( \frac{\partial
U_{B}}{\partial \psi} \right)^{\!\! -1} d \psi \Bigg| < + \infty .
\end{eqnarray}
Now, notice that the integral in (\ref{eq3: integr-div}) will
diverge either if $\partial_{\psi} U_{B}(\psi_{\infty})
\rightarrow 0$ or $U_{B}(\psi_{\infty}) \rightarrow \pm \infty$.
We have already studied the first case, which necessarily happens
at infinity. The second case, $U_{B}(\psi_{\infty}) \rightarrow
\pm \infty$, will be associated with singularities of the type
$\omega \rightarrow 0$. For example, the only way to approach a
singularity of the type $U_{B}(\psi) \rightarrow \pm \infty$, with
$z \rightarrow z_{\infty}^{-}$, that is from the left, will be
with $\partial_{\psi} U_{B} > 0$. This corresponds to $U_{B}(\psi)
\rightarrow + \infty$, and therefore to a warp factor going as
$\omega \rightarrow 0$ (the same argument follows for $z
\rightarrow z_{\infty}^{+}$). This means that the only type of
singularity, at a finite location in the bulk, will be of the type
$\omega \rightarrow 0$.

Summarizing, we have seen how the bulk geometry of the vacuum
static system is completely determined by the behaviours of $\psi$
and $\omega$. $\psi$ is a monotonic function of $z$, while
$\omega$ can increase or decrease depending on the sign of
$U_{B}$. The only singularities possible in the bulk are of the
form $\omega \rightarrow 0$. In general, this is a good reason to
consider two-brane models in the context of BPS-systems: the
singularity can be shielded from the first brane, and made to
disappear from the bulk, with the presence of the second brane.
However, the second brane can be
attracted towards the singularity and eventually hit it.

\subsection{A Few Examples}

\begin{figure}[ht]
\begin{center}
\includegraphics[width=0.4\textwidth]{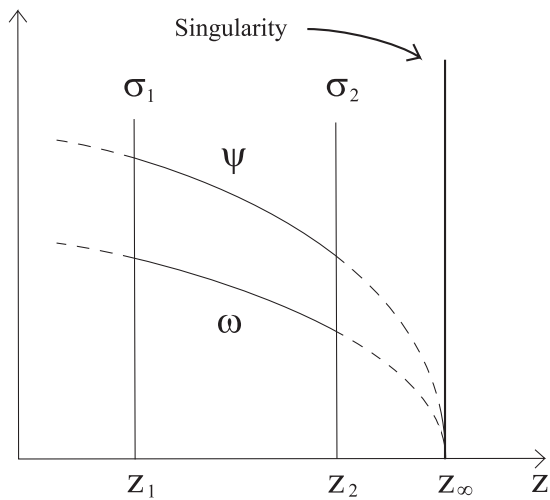}
\caption[Exponential Case]{The figure shows the solutions of
$\psi(z)$ and $\omega(z)$ for the exponential case, when the gauge
$N$ constant is adopted.} \label{F3}
\end{center}
\end{figure}

To gain some experience with BPS-configurations, let us briefly
study the static vacuum solution for a few choices of the
potential $U_{B}$. In particular, it will be useful to see how the
Randall-Sundrum model arises as a particular case of the present
type of system. Let us start by analyzing the case $U_{B} = V_{0}
e^{\alpha \psi}$, where $\alpha$ is an arbitrary constant. As
already said in Section \ref{sec: Intro}, this form of the
potential is motivated by heterotic M-theory, and is also
predicted in five-dimensional supergravity. Let us first consider
the gauge $N > 0$ a constant, and, for simplicity,
choose the positions of the branes at $z_{1} = 0$ and $z_{2} = r$,
with $r > 0$. Then, it follows that the solution to the system is
given by:
\begin{eqnarray}
\psi(z) &=&  - \frac{1}{\alpha} \ln \left[ - (N z + s) \, \alpha^{2} V_{0} \right], \\
\omega(z) &=&  \left[ - (N z + s) \, \alpha^{2} V_{0}
\right]^{\frac{1}{4 \alpha^{2}}}  \exp \left( \psi^{1} / 4 \alpha
\right),
\end{eqnarray}
where $N$ and $s$ can be expressed as:
\begin{eqnarray}
N &=&  - \frac{1}{r \alpha^{2} V_{0}} \left( e^{-\alpha \psi^{2}}
- e^{-\alpha \psi^{1}} \right), \\
s &=& - \frac{1}{\alpha^{2} V_{0}} e^{-\alpha \psi^{1}} .
\end{eqnarray}
For the singularities, there are two generic cases. When
$V_{0}>0$, the first brane is a positive tension brane and the
second one a negative tension brane, and a singularity of the type
$\omega = + \infty$ will exist at $z = -\infty$, while a
singularity of the type $\omega = 0$ will exist at the finite
position $z_{\infty} = - s / N = r \, e^{-\alpha \psi^{1}} / \big(
e^{-\alpha \psi^{1}} - e^{-\alpha \psi^{2}} \big) > r$. Thus, it
is not difficult to see that if $\alpha \psi^{2} \rightarrow +
\infty$, then the second brane hits the singularity $\omega = 0$.
Figure \ref{F3} shows the solutions for $\psi(z)$ and $\omega(z)$
in the present case (with the choice $\alpha < 0$). In the case
$V_{0}<0$, the first brane corresponds to a negative tension brane
and the second brane to a positive tension brane, with
singularities happening at the same coordinates but with opposite
signs.

The Randall-Sundrum scenario is easily obtained by letting $\alpha
\rightarrow 0$. In this limit, the above solutions can be
reexpressed as:
\begin{eqnarray}
\psi(z) &=&  0, \\
\omega(z) &=&  e^{- \mu N z},
\end{eqnarray}
where $\mu = (1/4) U_{B}$. In this case, the singularities are
located at infinity. Note that in this case the only degree of
freedom is $N$, the radion field.


\section{Low Energy Regime Expansion} \label{sec:
LER-expansion}

We are now in a position to deduce the equations of motion
governing the low energy regime. They consist in a linear
expansion of the fields about the static vacuum solution found in
the last section. As we shall see, these equations can be put in
an integral form, allowing the construction of a systematic scheme
to obtain solutions, order by order. The zeroth order solution of
this expansion corresponds to the static vacuum solutions with the
integration constants $\psi^{1}$ and $\psi^{2}$, and the metric
$\tilde g_{\mu \nu}$ promoted to be space-time dependent.

\subsection{Low Energy Regime Equations}

Let us develop the equations for the low energy regime. To start,
assume that $\psi_{0}$, $\omega_{0}$ and $N_{0}$ are functions of
$x$ and $z$ that satisfy the BPS-conditions:
\begin{eqnarray}
\frac{\omega_{0}'}{\omega_{0}} &=&
- \frac{1}{4} N_{0} U_{B}(\psi_{0}), \label{eq4: match-omega2} \\
\psi_{0}'  &=& N_{0} \frac{ \partial U_{B}}{\partial \psi_{0}},
\label{eq4: match-scalar2}
\end{eqnarray}
and let us define the bulk scalar field boundary values as
$\psi_{0}^{1} = \psi_{0}(z_{1})$ and $\psi_{0}^{2} =
\psi_{0}(z_{2})$ (note that they are functions of the space-time
coordinate $x$). The solution for the warp factor, $\omega_{0}$,
is found to be:
\begin{eqnarray}
\omega_{0} (z,x) = \exp \left[ -\frac{1}{4}
\int_{\psi_{0}^{1}}^{\psi_{0}} \alpha^{\! -1}(\psi) \, d\psi
\right],
\end{eqnarray}
where the $z$ and $x$ dependence enter through the functions
$\psi_{0}^{1}(x)$ and $\psi_{0}(z,x)$ in the integration limits.
It will be convenient to define the warp factor between the two
branes as $\bar \omega = \omega_{0}(z_{2})$, or more explicitly,
as:
\begin{eqnarray}
\bar \omega (x) = \exp \left[ -\frac{1}{4}
\int_{\psi_{0}^{1}}^{\psi_{0}^{2}} \alpha^{\! -1}(\psi) \, d\psi
\right].
\end{eqnarray}
Now, we would like to study the perturbed system about the static
vacuum solution. With this purpose, let us define the following
set of variables, $h_{\mu \nu}$, $\varphi$ and $\phi$, as:
\begin{eqnarray}
g_{\mu \nu} &=& \omega_{0}^{2} \left( \tilde g_{\mu \nu}
+ h_{\mu \nu} \right), \\
\psi &=& \psi_{0} + \varphi, \\
N &=& N_{0} \, e^{\phi},
\end{eqnarray}
where $g_{\mu \nu}$, $\psi$ and $N$ satisfy the equations of
motion (\ref{eq2: Grav-eq1}), (\ref{eq2: Grav-eq3}) and
(\ref{eq2: scalar}), taking into account the presence of matter in
the branes and supersymmetry breaking terms in the potentials.
Additionally, $\tilde g_{\mu \nu}$ depends only on the space-time
coordinate $x$. Note that if there is no matter content in the
branes and $u = v_{1} = v_{2} = 0$, then we can take $h_{\mu \nu}
= \varphi = \phi = 0$, and the fields $g_{\mu \nu}$, $\psi$ and
$N$ of the previous definition would correspond to the static vacuum
solution discussed in the last section, with $\psi_{0}^{1}$
and $\psi_{0}^{2}$ the two constant degrees of freedom. Therefore,
the functions $h_{\mu \nu}$, $\varphi$ and $\phi$ are linear
deviations from the vacuum solution of the system. Since we are
interested in the low energy regime, we consider the case $|h_{\mu
\nu}| \ll | \tilde g_{\mu \nu}|$, $|\varphi| \ll |\psi_{0}|$ and
$|\phi| \ll 1$. Now, if we insert these previous definitions in
the equations of motion (\ref{eq2: Grav-eq1}), (\ref{eq2:
Grav-eq3}) and (\ref{eq2: scalar}), and neglect second order
quantities in terms of $h_{\mu \nu}$, $\varphi$ and $\phi$ we
obtain the equations of motion for the low energy regime. First,
equation (\ref{eq2: Grav-eq1}) leads to:
\begin{eqnarray}
N_{0} U_{B}(\psi_{0}) \, \tilde g^{\alpha \beta} h_{\alpha \beta}'
+ 2 N_{0} \frac{\partial U_{B}}{\partial \psi_{0}} \varphi' \qquad
\quad \nonumber\\ - N^{2}_{0} \left[ 2 U_{0} \phi + \frac{\partial
U_{0}}{\partial \psi_{0}} \varphi \right] = \frac{N_{0}^{2}}{
\omega_{0}^{2}} \mathcal{X}. \label{eq4: lin-A}
\end{eqnarray}
Equation (\ref{eq2: Grav-eq3}) leads to the 4-dimensional
Einstein's equations:
\begin{eqnarray}
h_{\mu \nu}'' - \tilde g_{\mu \nu} \tilde g^{\alpha \beta}
h_{\alpha \beta}'' \nonumber\\ - \left[ N_{0} U_{B}(\psi_{0}) +
\frac{\partial N_{0}}{\partial \psi_{0}} \frac{\partial
U_{B}}{\partial \psi_{0}} \right] (h_{\mu \nu}' - \tilde g_{\mu
\nu} \tilde g ^{\alpha \beta} h_{\alpha \beta}') \nonumber\\ -
\frac{3}{2} N_{0} U_{B}(\psi_{0}) \, \phi' \, \tilde g_{\mu \nu} -
\frac{3}{2} N_{0} \frac{\partial U_{B}}{\partial \psi_{0}} \,
\varphi' \, \tilde g_{\mu \nu} \nonumber\\  - \frac{3}{4}
N_{0}^{2} \left[ 2 U_{0} \phi + \frac{\partial U_{0}}{\partial
\psi_{0}} \varphi \right] \tilde g_{\mu \nu} = 2
\frac{N_{0}^{2}}{\omega_{0}^{2}} \mathcal{Y}_{\mu \nu}.
\label{eq4: lin-B}
\end{eqnarray}
And finally, equation (\ref{eq2: scalar}) [with the help of
equation (\ref{eq2: Grav-eq1})] leads to:
\begin{eqnarray}
\varphi'' - \left[ N_{0} U_{B}(\psi_{0}) + \frac{\partial
N_{0}}{\partial \psi_{0}} \frac{\partial U_{B}}{\partial \psi_{0}}
\right] \varphi' \nonumber\\ - N_{0} \frac{\partial
U_{B}}{\partial \psi_{0}} \phi'  + \frac{1}{2} N_{0}
\frac{\partial U_{B}}{\partial \psi_{0}} \tilde g^{\mu \nu} h_{\mu
\nu}' \nonumber\\ - \frac{N_{0}^{2}}{2} \left[ 2 \phi
\frac{\partial U_{0}}{\partial \psi_{0}} + \frac{\partial^{2}
U_{0}}{\partial \psi_{0}^{2}} \varphi \right] = \frac{N^{2}_{0}}{
\omega_{0}^{2}} \left( \mathcal{Z} - \frac{1}{2} \alpha_{0}
\mathcal{X} \right). \label{eq4: lin-C}
\end{eqnarray}
In the previous equations, $U_{0} = U(\psi_{0})$. Equations
(\ref{eq4: lin-A}) and (\ref{eq4: lin-B}) correspond to the
linearized Einstein's equations, while equation (\ref{eq4: lin-C})
corresponds to the linearized bulk scalar field equation. In the
previous set of equations we have defined, for the sake of
notation, the functions $\mathcal{X}$, $\mathcal{Y}_{\mu \nu}$ and
$\mathcal{Z}$, in the following way:
\begin{eqnarray}
\mathcal{X} &=&  \omega_{0}^{2} \left[ (\partial \psi)^{2} -
\frac{4}{3} R +  u \right] ,
\\
\mathcal{Y}_{\mu \nu} &=&  G_{\mu \nu} + \frac{1}{N} \big( g_{\mu
\nu} \Box N - \nabla_{\mu}
\nabla_{\nu} N   \big) \nonumber\\
&& + \frac{3}{4} \bigg( \frac{1}{2} g_{\mu \nu} (\partial
\psi)^{2} -
\partial_{\mu} \psi \partial_{\nu} \psi + \frac{1}{2} g_{\mu \nu} u  \bigg) , \quad \\
\mathcal{Z}  &=&  \omega_{0}^{2} \left[ \frac{1}{2} \frac{\partial
u }{\partial \psi_{0}} - \frac{1}{N} g^{\mu \nu} \nabla_{\mu} (N
\partial_{\nu} \psi) \right] .
\end{eqnarray}
Note that in the functions $\mathcal{X}$, $\mathcal{Y}_{\mu \nu}$
and $\mathcal{Z}$, the quantities $g_{\mu \nu}$, $\psi$ and $N$
are not explicitly expanded. Their expansions can be considered in
the following way:
\begin{eqnarray}
\mathcal{X} &=& X_{0} + X, \nonumber\\
\mathcal{Y}_{\mu \nu} &=& Y_{\mu \nu}^{0} + Y_{\mu \nu}, \nonumber\\
\mathcal{Z} &=& Z_{0} + Z, \label{eq4: X-Y-Z}
\end{eqnarray}
where $X_{0}$, $Y_{\mu \nu}^{0}$ and $Z_{0}$ are the zeroth order
terms in the expansions of $\mathcal{X}$, $\mathcal{Y}_{\mu \nu}$
and $\mathcal{Z}$ (that is, they are constructed from $\tilde
g_{\mu \nu}$, $\psi_{0}$ and $N_{0}$) while $X$, $Y_{\mu \nu}$ and
$Z$ are the terms which contain linear contributions from  $h_{\mu
\nu}$, $\varphi$ and $\phi$. Their specific form are given in
Appendix \ref{app: Def}. We can now replace the new set of
variables in the Israel matching conditions and the scalar field
boundary condition. At the first brane, $\sigma_{1}$, these take
the form:
\begin{eqnarray}
h_{\mu \nu}' - \tilde g_{\mu \nu} \tilde g^{\alpha \beta}
h_{\alpha \beta}' = \frac{3}{2} N_{0} U_{B}(\psi_{0}) \, \phi \,
\tilde g_{\mu \nu} \qquad \nonumber\\ +  \frac{3}{2} N_{0}
\frac{\partial U_{B}}{\partial \psi_{0}} \, \varphi \, \tilde
g_{\mu \nu}  + \frac{3}{2} N_{0} v_{1} \tilde g_{\mu \nu} -
\kappa_{5}^{2} N_{0} T_{\mu \nu}^{\,1} , \label{eq4: lin-bound 1}
\end{eqnarray}
and
\begin{eqnarray}
\varphi' = N_{0} \frac{\partial U_{B}}{\partial \psi_{0}} \phi +
N_{0} \frac{\partial^{2} U_{B}}{\partial \psi_{0}^{2}} \varphi +
N_{0} \frac{
\partial v_{1}}{\partial \psi_{0}}, \label{eq4: lin-bound 2}
\end{eqnarray}
where, quantities like $\psi_{0}$ and $N_{0}$, not explicitly
evaluated at the boundary, must be evaluated at $z = z_{1}$.
Meanwhile, at the second brane, $\sigma_{2}$, the matching
conditions are:
\begin{eqnarray}
h_{\mu \nu}' - \tilde g_{\mu \nu} \tilde g^{\alpha \beta}
h_{\alpha \beta}' = \frac{3}{2} N_{0} U_{B}(\psi_{0}) \, \phi \,
\tilde g_{\mu \nu} \qquad \nonumber\\ +  \frac{3}{2} N_{0}
\frac{\partial U_{B}}{\partial \psi_{0}} \, \varphi \, \tilde
g_{\mu \nu}  + \frac{3}{2} N_{0} v_{2} \tilde g_{\mu \nu} +
\kappa_{5}^{2} N_{0} T_{\mu \nu}^{\, 2} \bar \omega_{0}^{-2} ,
\label{eq4: lin-bound 3}
\end{eqnarray}
and
\begin{eqnarray}
\varphi' = N_{0} \frac{\partial U_{B}}{\partial \psi_{0}} \phi +
N_{0} \frac{\partial^{2} U_{B}}{\partial \psi_{0}^{2}} \varphi +
N_{0} \frac{
\partial v_{2}}{\partial \psi_{0}}, \label{eq4: lin-bound 4}
\end{eqnarray}
where again, quantities like $\psi_{0}$ and $N_{0}$ must be
evaluated at $z = z_{2}$. Note that in the boundary conditions the
new set of variables are linearly proportional to the
energy-momentum tensor at the branes. In other words, deviations
of the boundary conditions from the BPS-state generate the existence of
fields $h_{\mu \nu}$, $\varphi$ and $\phi$ in the bulk, as
expected in the low energy regime. In order to solve the equations
of motion it is useful to rewrite them in the form of a set of
first order differential equations in terms of
the $\partial_{z}$ derivative.
This can be done by defining new fields $\xi_{\mu \nu}$ and $\chi$
to be sources of the fields  $h_{\mu \nu}$, $\varphi$ and $\phi$,
in the following equations:
\begin{eqnarray}
h_{\mu \nu}' - \tilde g_{\mu \nu} \tilde g ^{\alpha \beta}
h_{\alpha \beta}' - \frac{3}{2} N_{0} U_{B}(\psi_{0}) \, \phi \,
\tilde g_{\mu \nu} \nonumber\\ - \frac{3}{2} N_{0} \frac{\partial
U_{B}}{\partial \psi_{0}} \, \varphi \, \tilde
g_{\mu \nu} &=& \xi_{\mu \nu} , \\
\varphi' - N_{0} \frac{\partial U_{B}}{\partial \psi_{0}} \phi -
N_{0} \frac{\partial^{2} U_{B}}{\partial \psi_{0}^{2}} \varphi &=&
\chi.
\end{eqnarray}
With these definitions of $\xi_{\mu \nu}$ and $\chi$,
it is found that equations (\ref{eq4:
lin-A}), (\ref{eq4: lin-B}) and (\ref{eq4: lin-C}) can be
reexpressed in a much simpler way. Respectively, these equations
are found to be:
\begin{eqnarray}
2 N_{0} \frac{\partial U_{B}}{\partial \psi_{0}} \chi -
\frac{1}{3} N_{0} U_{B}(\psi_{0}) \xi =
\frac{N_{0}^{2}}{\omega_{0}^{2}} \mathcal{X}, \label{eq4:
R-linear}
\\
\left[ \frac{\omega_{0}^{4}}{N_{0}} \, \xi_{\mu \nu} \right]'  = 2
N_{0} \omega_{0}^{2} \mathcal{Y}_{\mu \nu}, \label{eq4:
Ein-linear}
\\
\left[ \alpha_{0} \frac{\omega_{0}^{4}}{N_{0}} \chi \right]' =
\alpha_{0} N_{0} \omega_{0}^{2} \left( \mathcal{Z} - \frac{1}{2}
\alpha_{0} \mathcal{X} \right) , \label{eq4: Sca-linear}
\end{eqnarray}
where $\xi = \tilde g^{\mu \nu} \xi_{\mu \nu}$. Note that the left
hand sides of equations (\ref{eq4: Ein-linear}) and (\ref{eq4:
Sca-linear}) consist of first order derivatives of $\xi_{\mu \nu}$
and $\chi$ in terms of $z$, while the right hand sides depends on second
order derivatives in terms of the space-time coordinate $x$.
The boundary values of $\xi_{\mu \nu}$ and $\chi$, at the
positions of the branes, can be
computed with the help of the boundary conditions (\ref{eq4:
lin-bound 1})-(\ref{eq4: lin-bound 4}). These are:
\begin{eqnarray}
\xi_{\mu \nu}(z_{1}) &=& \frac{3}{2} N_{0} v_{1} \tilde g_{\mu
\nu} - \kappa_{5}^{2} N_{0} T_{\mu \nu}^{\,1}, \label{eq4: bound1} \\
\chi(z_{1})  &=& N_{0}  \frac{
\partial v_{1}}{\partial \psi_{0}}, \label{eq4: bound2}
\end{eqnarray}
at position $z_{1}$, and
\begin{eqnarray}
\xi_{\mu \nu}(z_{2}) &=& \frac{3}{2} N_{0} v_{2} \tilde g_{\mu
\nu} + \kappa_{5}^{2} N_{0} T_{\mu \nu}^{\,2} \bar \omega^{-2},
\label{eq4: bound3} \\
\chi(z_{2})  &=& N_{0}  \frac{
\partial v_{2}}{\partial \psi_{0}}, \label{eq4: bound4}
\end{eqnarray}
at position $z_{2}$. (Recall that $\omega_{0}(z_{1}) = 1$ and
$\omega_{0}(z_{2}) = \bar \omega$). We should not forget at this
point the additional equations for the matter content at both
branes. They come from equation (\ref{eq2: conserv}) and, with the present
notation, are given by:
\begin{eqnarray}
\tilde \nabla^{\nu} T^{1}_{\mu \nu} &=& 0, \label{eq4: cons-T1}\\
\tilde \nabla^{\nu} T^{2}_{\mu \nu} &=& \frac{1}{\bar \omega}
\left[ (\tilde \nabla_{\mu} \bar \omega) \tilde g^{\alpha \beta}
T^{2}_{\alpha \beta} - 2 (\tilde \nabla^{\nu} \bar \omega)
T^{2}_{\mu \nu} \right] \label{eq4: cons-T2},
\end{eqnarray}
where covariant derivatives $\tilde \nabla_{\mu}$ are constructed
from the $\tilde g_{\mu \nu}$ metric. Additionally, it is worth
mentioning that the projected Weyl tensor $E_{\mu \nu}$ and the
loss parameter $\Delta \psi$ can be written in terms of the linear
variables as:
\begin{eqnarray}
E_{\mu \nu} &=& -\frac{\omega_{0}^{2}}{2 N_{0}} \left[ \xi_{\mu
\nu}' - \frac{1}{3} \tilde g_{\mu \nu} \tilde g^{\alpha \beta}
\xi_{\alpha \beta}' \right] \nonumber\\ && + \frac{1}{4}
\frac{\omega_{0}^{2}}{N_{0}} U_{B}(\psi_{0}) \left[ \xi_{\mu \nu}
- \frac{1}{3} \tilde g_{\mu \nu} \tilde g^{\alpha \beta}
\xi_{\alpha \beta} \right] \nonumber\\
&& - \frac{3}{8} \tilde g_{\mu \nu} \frac{\partial U_{B}}{\partial
\psi_{0} } \frac{\chi}{N_{0}} -  \frac{1}{4} \bigg[ \frac{1}{4}
g_{\mu \nu} u + \partial_{\mu} \psi
\partial_{\nu} \psi \nonumber\\ && - \frac{1}{4} g_{\mu
\nu} (\partial \psi)^{2} +
\frac{4}{N} \nabla_{\mu} \nabla_{\nu} N \bigg], \label{eq4: E-Weyl}\\
\Delta \psi &=& \frac{1}{N_{0}} \left[ \frac{\chi}{N_{0}}
\right]'. \label{eq4: Delta-psi}
\end{eqnarray}

One of the main features of equations (\ref{eq4: Ein-linear}) and
(\ref{eq4: Sca-linear}) is that now they can be put in an integral
form. That is, we can write:
\begin{eqnarray}
\frac{\omega_{0}^{4}}{N_{0}} \xi_{\mu \nu}(z) &=&  \frac{3}{2}
\tilde g_{\mu \nu} v_{1}  - \kappa_{5}^{2} T_{\mu \nu}^{\,1}
\nonumber\\ && + 2
\int_{z_{1}}^{z} dz \, N_{0} \omega_{0}^{2} \mathcal{Y}_{\mu \nu}, \label{eq4: inte-1}\\
\alpha_{0} \frac{\omega_{0}^{4}}{N_{0}} \chi(z) &=& \alpha_{0}
(\psi_{0}^{1}) \frac{
\partial v_{1}}{\partial \psi_{0}^{1}} \nonumber\\ && +
\int_{z_{1}}^{z} dz \, \alpha_{0} N_{0} \omega_{0}^{2} \left(
\mathcal{Z} - \frac{1}{2} \alpha_{0} \mathcal{X} \right),
\label{eq4: inte-2}
\end{eqnarray}
where we have used the boundary conditions at the position of the
first brane. To further proceed we must implement a systematic
expansion order by order in which the vacuum solution serves as
the zeroth order solution. We develop this in the next
subsection.

\subsection{Low Energy Regime Expansion}

Equations (\ref{eq4: inte-1}) and (\ref{eq4: inte-2}) are two
integral equations of the system, which suggest the possibility of
introducing a systematic expansion about the vacuum solution,
order by order. In the last subsection we saw that the matter
content of the branes are sources for the linear deviations $h_{\mu
\nu}$, $\varphi$ and $\phi$ on the bulk. Therefore, it is sensible
to study how the bulk is affected by the matter distribution on
the brane, at different scales. In this way, let us consider the
following expansion for the energy-momentum tensor of the matter
content on the branes, as well as for the supersymmetry breaking
potentials:
\begin{eqnarray}
T_{\mu \nu}^{1} &=& T_{\mu \nu}^{1 \, (0)} + T_{\mu \nu}^{1 \,
(1)}
+ T_{\mu \nu}^{1 \, (2)} + \cdots, \\
v_{1} &=& v_{1}^{(0)} + v_{1}^{(1)} + v_{1}^{(2)} + \cdots ,
\end{eqnarray}
and
\begin{eqnarray}
T_{\mu \nu}^{2} &=& T_{\mu \nu}^{2 \, (0)} + T_{\mu \nu}^{2 \,
(1)}
+ T_{\mu \nu}^{2 \, (2)} + \cdots, \\
v_{2} &=& v_{2}^{(0)} + v_{2}^{(1)} + v_{2}^{(2)} + \cdots .
\end{eqnarray}
The parameter of the expansion is dictated by the scale at which
each term becomes relevant in the physical problem of interest.
Naturally, the expansion above will
induce an expansion of the
source functions $\xi_{\mu \nu}$ and $\chi$, given by:
\begin{eqnarray}
\xi_{\mu \nu} &=& \xi_{\mu \nu}^{(1)} + \xi_{\mu \nu}^{(2)} + \cdots, \\
\chi &=& \chi_{(1)} + \chi_{(2)} + \cdots,
\end{eqnarray}
in this way, the boundary conditions
(\ref{eq4: bound1})-(\ref{eq4: bound4}), can now be
written, order by order, as:
\begin{eqnarray}
\xi_{\mu \nu}^{(i+1)}(z_{1}) &=& \frac{3}{2} N_{0} v_{1}^{(i)}
\tilde g_{\mu
\nu} - \kappa_{5}^{2} N_{0} T_{\mu \nu}^{1 \, (i)}, \\
\chi_{(i + 1)}(z_{1})  &=& N_{0}  \frac{
\partial v_{1}^{(i)}}{\partial \psi_{0}},
\end{eqnarray}
at position $z_{1}$, and
\begin{eqnarray}
\xi_{\mu \nu}^{(i+1)}(z_{2}) &=& \frac{3}{2} N_{0} v_{2}^{(i)}
\tilde g_{\mu
\nu} + \kappa_{5}^{2} N_{0} T_{\mu \nu}^{2 \, (i)} \bar \omega^{-2}, \\
\chi_{(i + 1)}(z_{2})  &=& N_{0}  \frac{
\partial v_{2}^{(i)}}{\partial \psi_{0}},
\end{eqnarray}
at position $z_{2}$. Note the convention whereby the indicies of the left
hand side are raised by one unit in relation those on the right hand side.
Continuing with the construction, we must also consider the expansion
of $h_{\mu\nu}$, $\varphi$ and $\phi$:
\begin{eqnarray}
h_{\mu \nu} &=& h_{\mu \nu}^{( 1)} + h_{\mu \nu}^{(2)} + \cdots, \label{eq4: h-exp} \\
\varphi &=& \varphi_{(1)} + \varphi_{(2)} + \cdots, \label{eq4: psi-exp}\\
\phi &=& \phi_{(1)} + \phi_{(2)} + \cdots. \label{eq4: N-exp}
\end{eqnarray}
They are defined to satisfy the following first order differential
equations with sources $\xi_{\mu \nu}^{(i)}$ and $\chi_{(i)}$:
\begin{eqnarray}
h_{\mu \nu}^{(i)} \,\! ' - \tilde g_{\mu \nu} \tilde g^{\alpha
\beta} h_{\alpha \beta}^{(i)} \,\! ' - \frac{3}{2} N_{0}
U_{B}(\psi_{0}) \, \phi_{(i)} \, \tilde g_{\mu \nu} \nonumber\\
- \frac{3}{2} N_{0} \frac{\partial U_{B}}{\partial \psi_{0}} \,
\varphi_{(i)} \, \tilde g_{\mu \nu} = \xi_{\mu \nu}^{(i)},
 \label{eq4: def xi-i}\\
\varphi_{(i)}' - N_{0} \frac{\partial U_{B}}{\partial \psi_{0}}
\phi_{(i)} - N_{0} \frac{\partial^{2} U_{B}}{\partial
\psi_{0}^{2}} \varphi_{(i)} = \chi_{(i)}. \label{eq4: def chi-i}
\end{eqnarray}
Finally, recall that the quantities $\mathcal{X}$,
$\mathcal{Y}_{\mu \nu}$ and $\mathcal{Z}$ defined in equations
(\ref{eq4: X-Y-Z}) depend on the fields $h_{\mu \nu}$, $\varphi$ and $\phi$.
Therefore, we must consider the following expansion
\begin{eqnarray}
\mathcal{X} &=& X_{0} + X_{(1)} + X_{(2)} + \cdots, \label{eq4: A-exp}\\
\mathcal{Y}_{\mu \nu} &=& Y_{\mu \nu}^{0} + Y_{\mu \nu}^{(1)} +
Y_{\mu
\nu}^{(2)} + \cdots, \label{eq4: B-exp} \\
\mathcal{Z} &=& Z_{0} + Z_{(1)} + Z_{(2)} + \cdots, \label{eq4:
C-exp}
\end{eqnarray}
where the index ``$0$'' denotes the dependence on the zeroth order
quantities, $N_{0}$, $\psi_{0}$ and $\omega_{0}$, and the index $i
\geq 1$ denotes quantities that depend linearly on $h_{\mu
\nu}^{(i)}$, $\varphi_{(i)}$ and $\phi_{(i)}$. The precise form of
the expanded functions $X_{(i)}$, $Y^{(i)}_{\mu \nu}$ and
$Z_{(i)}$ are shown in Appendix \ref{app: Def}. Observe that it
follows from equations (\ref{eq4: cons-T1}) and (\ref{eq4:
cons-T2}) that the $i$-th order term, $T_{\mu \nu}^{a \, (i)}$, in
the expansion of the energy momentum tensor for the matter living
in the brane, satisfies the following conservation relation:
\begin{eqnarray}
\tilde \nabla^{\nu} T^{1 \,(i)}_{\mu \nu} &=& 0, \\
\tilde \nabla^{\nu} T^{2 \, (i)}_{\mu \nu} &=& \frac{1}{\bar \omega} \big[
(\tilde \nabla_{\mu} \bar \omega) \tilde g^{\alpha \beta}
T^{2 \, (i)}_{\alpha \beta}
\nonumber\\ && - 2 (\tilde \nabla^{\nu} \bar \omega)
T^{2 \, (i)}_{\mu \nu} \big].
\end{eqnarray}

With all these previous definitions we can now cast the equations
of motion in the following way:
\begin{eqnarray}
2 \frac{\partial U_{B}}{\partial \psi_{0}} \chi_{(i+1)} -
\frac{1}{3} U_{B}(\psi_{0}) \xi_{(i+1)} =
\frac{N_{0}}{\omega_{0}^{2}} X_{(i)},
\label{eq4: R-constr} \\
\left[ \frac{\omega_{0}^{4}}{N_{0}} \xi_{\mu \nu}^{(i+1)} \right]'
= 2 N_{0} \omega_{0}^{2} Y^{(i)}_{\mu \nu}, \label{eq4: B-exp 2} \\
\left[ \alpha_{0} \, \frac{\omega_{0}^{4}}{N_{0}} \, \chi_{(i+1)}
\right]' = \alpha_{0} N_{0} \omega_{0}^{2} \left( Z_{(i)} -
\frac{1}{2} \alpha_{0} X_{(i)} \right). \label{eq4: A-C-exp}
\end{eqnarray}
This last set of equations shows the desired low energy regime
expansion. It states that $\xi_{\mu \nu}^{(i+1)}$ and
$\chi_{(i+1)}$ can be solved in terms of the lower order
quantities $\mathcal{X}_{(i)}$, $\mathcal{Y}^{(i)}_{\mu \nu}$ and
$\mathcal{Z}_{(i)}$. At the same time, the functions
$\mathcal{X}_{(i)}$, $\mathcal{Y}^{(i)}_{\mu \nu}$ and
$\mathcal{Z}_{(i)}$ can be solved in terms of $\xi_{\mu
\nu}^{(i)}$ and $\chi_{(i)}$, as evident from their definitions.
Note that the former means that we can compute $E_{\mu \nu}$ and
$\Delta \psi$ to any desired order, as indicated by
equations (\ref{eq4: E-Weyl}) and (\ref{eq4: Delta-psi}).
In this way, starting with the zeroth order solutions $\psi_{0}$,
$N_{0}$ and $\tilde g_{\mu \nu}$, we can arrive at any desired
order for the full solutions to the bulk equations
$\psi$, $N$ and $g_{\mu \nu}$. The present method is similar
to the one introduced by Kanno and Soda for the Randall-Sundrum
model \cite{Kanno-Soda 1, Kanno-Soda 2} as well as to other
schemes \cite{Kobayashi-Koyama, Mukohyama-Kofman, Mukohyama-Coley, Kanno-Soda 3}
for dilatonic brane-worlds.

We mentioned that the expansion parameter was dictated by the scale
at which each term of the energy-momentum tensor expansion was
relevant for the physical process of interest. In terms of the
bulk quantities, one finds that the effect on the variation of the
scale in the extra-dimensional direction is related to the
variation of the scales in the space-time direction as follows:
\begin{eqnarray}
h_{(i+1)}'' \simeq \frac{N_{0}^{2}}{\omega_{0}^{2}} \Box h_{(i)},
\label{eq4: exp-order-i}
\end{eqnarray}
That is, second derivatives of the $(i+1)$-th order variables in
terms of $z$ are of the same order as the second order derivatives
of the $i$-th order variables in terms of space-time coordinates.

As in the previous case, we can rewrite equations (\ref{eq4: B-exp
2}) and (\ref{eq4: A-C-exp}) in an integral form:
\begin{eqnarray}
\lefteqn{ \frac{\omega_{0}^{4}}{N_{0}} \xi^{(i+1)}_{\mu \nu} =
\frac{3}{2} \tilde g_{\mu \nu}  v_{1}  - \kappa_{5}^{2} T_{\mu
\nu}^{\,1} }  \nonumber\\ && {}
 + 2 \int_{z_{1}}^{z} \!\!\! dz
 N_{0} \omega_{0}^{2} Y^{(i)}_{\mu \nu}, \label{eq4: inte2-1}
\\
\lefteqn{ \alpha_{0} \frac{\omega_{0}^{4}}{N_{0}} \chi_{(i+1)} =
\alpha_{1} \frac{ \partial v_{1}}{\partial \psi_{0}^{1}} }
 \nonumber\\ && {} + \int_{z_{1}}^{z} \!\!\! dz  \alpha_{0}
N_{0} \omega_{0}^{2} \left( Z_{(i)} - \frac{1}{2} \alpha_{0}
X_{(i)} \right). \label{eq4: inte2-2}
\end{eqnarray}
This form of the expanded equations of motion are highly relevant;
they are the basis for the derivation of the effective theory
developed in the next section.

Summarizing, in this section we have developed an expansion
procedure to solve the entire theory in a systematic form. As we
said, to solve this system of equations we need to know the form
of the zeroth order moduli fields $\psi^{1}_{0}$ and $\psi^{2}_{0}$, and
the metric $\tilde g_{\mu \nu}$ and, therefore, an effective
theory for these fields is required. In the following section, we
are going to deduce the 4-dimensional equations of motion
governing these functions, as well as those for higher order
terms in the expansion.


\section{Four Dimensional Effective Theory} \label{sec: LER-eff}

In the previous section we have shown how to construct the low
energy regime as a consistent expansion. In this section we see
how to define the four-dimensional effective theory at the
position of the branes starting from the expansion above.
Concordant with the expansion, the effective theory must be
defined order by order.

\subsection{General Case}

The effective equations governing the variables at the
$i$-th order can be obtained by evaluating the integral equations
(\ref{eq4: inte2-1}) and (\ref{eq4: inte2-2}) at $z = z_{2}$. In
other words, the $i$-th order equations are given by:
\begin{eqnarray}
2 \int_{z_{1}}^{z_{2}} \!\!\! dz \, N_{0} \omega_{0}^{2}
Y^{(i)}_{\mu \nu} = \frac{3}{2} \big[ \bar \omega^{4} v_{2}^{(i)}
- v_{1}^{(i)} \big]
\tilde g_{\mu \nu} \nonumber\\
+ \kappa_{5}^{2} \left[ T_{\mu \nu}^{2 \, (i)} \bar \omega^{2} +
T_{\mu \nu}^{1 \, (i)} \right],
\label{eq5: eff-i-1} \\
\int_{z_{1}}^{z_{2}} \!\!\! dz \, \alpha_{0} N_{0} \omega_{0}^{2}
\bigg( Z_{(i)} - \frac{1}{2} \alpha_{0} X_{(i)} \bigg) =
\alpha_{2} \bar \omega^{4} \frac{
\partial v_{2}^{(i)} }{\partial \psi_{0}^{2}} \nonumber\\ - \alpha_{1}
\frac{ \partial v_{1}^{(i)} }{\partial \psi_{0}^{1}}, \label{eq5:
eff-i-2}
\end{eqnarray}
Additionally, another equation can be obtained by evaluating
equation (\ref{eq4: R-constr}) at the boundary position $z_{1}$.
The resulting equation is:
\begin{eqnarray}
X_{(i)}(z_{1}) &=& 2 \frac{\partial U_{B}}{\partial \psi_{0}^{1}}
\frac{
\partial v_{1}^{(i)}}{\partial \psi_{0}^{1}}  - 2 U_{B}(\psi_{0}^{1}) v_{1}^{(i)}
\nonumber\\ && + \frac{\kappa_{5}^{2}}{3}
 U_{B}(\psi_{0}^{1}) \,  \tilde g^{\mu \nu} T^{1 \, (i)}_{\mu \nu} .
\label{eq5: eff-i-3}
\end{eqnarray}
Note that a similar equation would have been obtained by
evaluating (\ref{eq4: R-constr}) at the boundary position $z_{2}$,
however, this would not give a new equation of motion. Equations
(\ref{eq5: eff-i-1})-(\ref{eq5: eff-i-3}) are the desired
equations of the 4-dimensional effective theory. Solving these
equations for the $i$-th order variables allows us to obtain the
effective equations for the $(i+1)$-th order variables, after
correctly integrating (\ref{eq4: def xi-i}) and (\ref{eq4: def
chi-i}). The form in which equations (\ref{eq5:
eff-i-1})-(\ref{eq5: eff-i-3}) are presented is, at this level,
abstract and it is difficult to appreciate the effective theory in
more familiar terms. In the next subsection we
analyze in detail the effective theory at the zeroth order in the
expansion ($i=0$).

\subsection{Zeroth Order Effective Theory}

It will be particularly interesting to analyze the form of the
effective equations at the zeroth order in more detail. In this
case the metric at the first and second branes are conformally
related. In terms of the treatment developed in the previous
section, the effective theory is:
\begin{eqnarray}
\int_{z_{1}}^{z_{2}} \!\!\!\!\! dz \, N_{0} \omega_{0}^{2}
Y^{0}_{\mu \nu} = \frac{3}{4} \tilde g_{\mu \nu} \left[ \bar
\omega^{4} v_{2}^{0}
- v_{1}^{(0)} \right] \nonumber\\
+ \frac{\kappa_{5}^{2}}{2} \left[ T_{\mu \nu}^{2 \, (0)} \bar
\omega^{2} + T_{\mu \nu}^{1 \, (0)} \right],
\label{eq5: eff-0-1} \\
\int_{z_{1}}^{z_{2}} \!\!\!\!\! dz \, \alpha_{0} N_{0}
\omega_{0}^{2} \bigg( Z_{0} - \frac{1}{2} \alpha_{0} X_{0} \bigg)
= \alpha_{2} \bar \omega^{4} \frac{
\partial v_{2}^{(0)} }{\partial \psi^{2}_{0}} \nonumber\\ - \alpha_{1} \frac{
\partial v_{1}^{(0)} }{\partial \psi^{1}_{0}}, \label{eq5: eff-0-2} \\
X_{0}(z_{1}) = 2 \frac{\partial U_{B}}{\partial \psi^{1}_{0}}  \frac{
\partial v_{1}^{(0)}}{\partial \psi^{1}_{0}}  - 2 U_{B}(\psi^{1}_{0}) v_{1}^{(0)}
\nonumber\\  + \frac{\kappa_{5}^{2}}{3}
 U_{B}(\psi^{1}_{0}) \, \tilde g^{\mu \nu}  T^{1 \, (0)}_{\mu \nu} . \label{eq5: eff-0-3}
\end{eqnarray}
In the following, we shall omit the ``$0$'' index denoting the
zeroth order terms. To rewrite this theory in more familiar terms,
we need to integrate the terms on the left hand side of equations
(\ref{eq5: eff-0-1}) and (\ref{eq5: eff-0-2}). In order to do this
we need to evaluate the terms at the boundaries. Hence it is
particularly useful to express the theory in terms of the moduli
fields $\psi^{1}$ and $\psi^{2}$. It is then possible to obtain
the following effective theory (Appendix \ref{app: Comp} shows how
to derive the next results). The Einstein's equations, obtained
from (\ref{eq5: eff-0-1}), are found to be:
\begin{eqnarray}
\Omega^{2} \tilde G_{\mu \nu} = - \tilde g_{\mu \nu} \tilde \Box \, \Omega^{2}
+ \tilde \nabla_{\mu} \tilde \nabla_{\nu} \Omega^{2}  \qquad \qquad
\qquad \nonumber\\  - \frac{3}{4}
\gamma_{a b} \bigg( \frac{1}{2} \tilde g_{\mu \nu}  \tilde g^{\alpha
\beta} \partial_{\alpha} \psi^{a}  \partial_{\beta} \psi^{b} -
\partial_{\mu} \psi^{a}  \partial_{\nu} \psi^{b} \bigg) \nonumber\\  +
\frac{k \kappa_{5}^{2}}{2}  \left[ \bar \omega^{2}  T_{\mu
\nu}^{2} + T_{\mu \nu}^{1} \right] - \frac{3}{4} \tilde g_{\mu
\nu} V, \label{eq5: EFF1}
\end{eqnarray}
while the  moduli fields equations, obtained from equations
(\ref{eq5: eff-0-1}) and (\ref{eq5: eff-0-3}), are found to be:
\begin{eqnarray}
\tilde \Box \psi^{a} &=& - \tilde g^{\mu \nu} \gamma^{a}_{b c}
\partial_{\mu} \psi^{b} \partial_{\nu} \psi^{c}
+ \frac{1}{2} \gamma^{a c} \frac{\partial V}{\partial \psi^{c}}\nonumber\\
&& - \frac{2}{3} \gamma^{a c} \frac{\partial \Omega^{2}}{\partial
\psi^{c}} \tilde R   - \frac{2}{3} k \kappa_{5}^{2} \bar \omega
\gamma^{a c} \frac{\partial \bar \omega}{\partial \psi^{c}} \tilde
g^{\mu \nu} T^{2}_{\mu \nu}. \qquad \label{eq5: EFF2}
\end{eqnarray}
In the previous expressions, the index $a$ labels the
positions $1$ and $2$. Additionally, note the presence of a
conformal factor $\Omega^{2}$ in front of the Einstein's tensor
$G_{\mu \nu}$, given by:
\begin{eqnarray}
\Omega^{2} = k \int_{\psi^{1}}^{\psi^{2}} \!\!\!\!\! d\psi \left(
\frac{\partial U_{B}}{\partial \psi} \right)^{-1} \!\!\!
\omega^{2}, \label{eq5: Big-Omega}
\end{eqnarray}
where $\omega$ is given by equation (\ref{eq3: omega}). The
coefficient $k$ is an arbitrary positive constant with dimensions
of inverse length, which has been incorporated to make
$\Omega^{2}$ dimensionless. The symmetric matrix $\gamma_{a b}$ is
a function of the moduli fields, that can be regarded as the
metric of the space spanned by the moduli in a sigma model
approach, with $\gamma^{a b}$ its inverse. The elements of
$\gamma_{a b}$ are given by:
\begin{eqnarray}
\gamma_{1 1} &=& \alpha_{1}^{-2} \left[ \frac{k}{U_{B}(\psi^{1})}
- \frac{1}{2} \Omega^{2}  \right],  \\
\gamma_{2 2} &=& \alpha_{2}^{-2}  \frac{\bar
\omega^{2} k}{U_{B}(\psi^{2})}, \\
\gamma_{1 2} &=& - \alpha_{1}^{-1} \alpha_{2}^{-1} \frac{\bar
\omega^{2} k}{U_{B}(\psi^{2})},
\end{eqnarray}
with $\gamma_{2 1} = \gamma_{1 2}$. Associated with the above
metric, we have defined a set of connections $\gamma^{a}_{bc}$.
These are given by:
\begin{eqnarray}
\gamma^{a}_{bc} = \frac{1}{2} \gamma^{a d} \left[ \frac{\partial
\gamma_{b d}}{\partial \psi^{c}} + \frac{\partial \gamma_{d
c}}{\partial \psi^{b}} - \frac{\partial \gamma_{b c}}{\partial
\psi^{d}} \right].
\end{eqnarray}
Finally, we have also defined an effective potential $V$ which
depends linearly on the supersymmetry breaking potentials $u$,
$v_{1}$ and $v_{2}$. This is defined as:
\begin{eqnarray}
V = \frac{k}{2} \int^{\psi^{2}}_{\psi^{1}} \!\!\!\!\! d\psi \left(
\frac{\partial U_{B}}{\partial \psi} \right)^{-1} \!\!\!
\omega^{4} \, u - 2 k \left[ \bar \omega^{4} v_{2} - v_{1}
\right].
\end{eqnarray}
Finally, we can not forget the matter conservation relations.
These take the form:
\begin{eqnarray}
\tilde \nabla^{\nu} T^{1}_{\mu \nu} &=& 0, \\
\tilde \nabla^{\nu} T^{2}_{\mu \nu} &=& \frac{1}{\bar \omega}
\big[ (\tilde \nabla_{\mu} \bar \omega) \tilde g^{\alpha \beta}
T^{2}_{\alpha \beta}  - 2 (\tilde \nabla^{\nu} \bar \omega)
T^{2}_{\mu \nu} \big].
\end{eqnarray}
The fact that the energy-momentum tensor of the second brane does
not respect the standard matter conservation relation is due to
the frame chosen to describe the system.

The generic form of the theory displayed by equations (\ref{eq5:
EFF1}) and (\ref{eq5: EFF2}) is of a bi-scalar tensor theory of
gravity, with the two scalar degrees given by $\psi^{1}$ and
$\psi^{2}$. The above set of equations can be obtained from the
following action:
\begin{eqnarray}
S &=& \frac{1}{k \kappa_{5}^{2}} \int d^{4}x \sqrt{- \tilde g}
\bigg[ \Omega^{2} \tilde R - \frac{3}{4} \tilde g^{\mu \nu}
\gamma_{a b}
\partial_{\mu} \psi^{a} \partial_{\nu} \psi^{b}
- \frac{3}{4} V  \bigg]  \nonumber\\
&& + S_{1}[\Psi_{1}, \tilde g_{\mu \nu}] + S_{2}[\Psi_{2}, \bar
\omega^{2} \tilde g_{\mu \nu}], \qquad \label{eq5: EFF Action}
\end{eqnarray}
Equation (\ref{eq5: EFF Action}) is an important result. It can be
shown that the effective action (\ref{eq5: EFF Action})
corresponds to the moduli-space approximation \cite{prep}, where
the relevant fields of the theory are just simply the free degrees
of freedom of the vacuum theory, promoted to be space-time
dependent fields. A relevant aspect of this theory, is that the
equation of motion for a moduli field $\psi^{a}$ will depend
linearly on the trace of the energy-momentum tensor of the brane
$a$ but not on the one belonging to the opposite brane. More
precisely, it can be shown that (\ref{eq5: EFF2}) has the form:
\begin{eqnarray}
\tilde \Box \psi^{1} &=& - \frac{k \kappa_{5}}{6} \frac{\partial
U_{B}}{\partial \psi^{1}} \tilde g^{\mu \nu} T^{1}_{\mu \nu} +
\cdots, \label{eq5: box-psi1}
\\
\tilde \Box \psi^{2} &=& + \frac{k \kappa_{5}}{6} \frac{\partial
U_{B}}{\partial \psi^{2}} \bar \omega^{-2} \tilde g^{\mu \nu}
T^{2}_{\mu \nu} + \cdots. \label{eq5: box-psi2}
\end{eqnarray}
This means that, in the low energy regime, the moduli fields are
driven by the matter content of the branes (recall that the moduli
are parameterizing the positions of the branes). This behaviour of
the moduli fields is independent of the frame choice (see next
subsection).

\subsection{Einstein's Frame}

Note that in equation (\ref{eq5: EFF Action}) the Newton's
constant depends on the moduli fields. This theory can be
rewritten in the Einstein frame where the Newton's constant is
independent of the moduli. Considering the conformal
transformation:
\begin{eqnarray}
\tilde g_{\mu \nu} \rightarrow g_{\mu \nu} = \Omega^{2} \tilde g_{\mu \nu},
\end{eqnarray}
we are then left with the following action
\begin{eqnarray}
S &=& \frac{1}{k \kappa_{5}^{2}} \int d^{4}x \sqrt{- g} \bigg[ R -
\frac{3}{4} g^{\mu \nu} \gamma_{a b}
\partial_{\mu} \psi^{a} \partial_{\nu} \psi^{b} - \frac{3}{4} V \bigg]  \nonumber\\
&& + S_{1}[\Psi_{1}, A_{1}^{2} g_{\mu \nu}] + S_{2}[\Psi_{2},
A_{2}^{2} g_{\mu \nu}], \label{eq5: EFF Action 3}
\end{eqnarray}
where now the sigma model metric $\gamma_{a b}$ is given by:
\begin{eqnarray}
\gamma_{1 1} &=&  2 \alpha_{1}^{-2} \frac{k^{2} A_{1}^{4}}
{U_{B}^{2}(\psi^{1})} \bigg[ 1 - \frac{1}{2 k} U_{B}(\psi^{1})
A_{1}^{-2}
\bigg] , \\
\gamma_{2 2} &=& 2 \alpha_{2}^{-2} \frac{k^{2} A_{2}^{4}}
{U_{B}^{2}(\psi^{2})} \bigg[ 1 + \frac{1}{2 k} U_{B}(\psi^{2})
A_{2}^{-2}
\bigg]  , \\
\gamma_{1 2} &=& - 2 \alpha_{1} \alpha_{2} \frac{ k^{2} A_{1}^{2}
A_{2}^{2} } {U_{B}(\psi^{1})  U_{B}(\psi^{2}) }.
\end{eqnarray}
As usual, $\gamma^{a b}$ is the inverse of $\gamma_{a b}$, and
$\gamma^{c}_{a b}$ are the connections. It is possible to show
that $\gamma_{a b}$ is a positive definite metric. To work in this
frame is particularly useful and simple. Additionally, we have
defined the quantities $A_{1}$ and $A_{2}$ (which are functions of
the moduli) to be:
\begin{eqnarray}
A_{1}^{-2} &=& \Omega^{2} \nonumber\\
&=& k \!\! \int_{\psi^{1}}^{\psi^{2}} \!\!\!\!\! d\psi \left(
\frac{\partial U_{B}}{\partial \psi} \right)^{\!\! -1} \!\!\!\!\!
\exp \left[ -\frac{1}{2} \int_{\psi^{1}}^{\psi}
\!\!\!\! \alpha^{-1} d \psi \right], \qquad \\
A_{2}^{-2} &=& \Omega^{2} \bar \omega^{-2} \nonumber\\
&=& k \!\! \int_{\psi^{1}}^{\psi^{2}} \!\!\!\!\! d\psi \left(
\frac{\partial U_{B}}{\partial \psi} \right)^{\!\! -1} \!\!\!\!\!
\exp \left[ -\frac{1}{2} \int_{\psi^{2}}^{\psi} \!\!\!\!
\alpha^{-1} d \psi \right]. \qquad
\end{eqnarray}
Also, the potential $V$ is now found to be:
\begin{eqnarray}
V &=& \frac{k}{2} \Omega^{-4} \int^{\psi^{2}}_{\psi^{1}}
\!\!\!\!\! d\psi \left( \frac{\partial U_{B}}{\partial \psi}
\right)^{-1} \!\!\! \omega^{4} \, u \nonumber\\ && - 2 k
\left[ A_{2}^{4} v_{2} - A_{1}^{4} v_{1} \right]. \label{eq5: Pot
V}
\end{eqnarray}
The present form of the theory, can be further worked out to put the sigma
model in a diagonal form with the  redefinition of the moduli
fields. However, in the present work we shall continue with the
theory in the current notation. Let us finish this section by
expressing the equations of motion in the Einstein frame. The Einstein's equations are:
\begin{eqnarray}
G_{\mu \nu} &=& - \frac{3}{4} \gamma_{a b} \bigg( \frac{1}{2}
 g_{\mu \nu}  g^{\alpha \beta}
\partial_{\alpha} \psi^{a}  \partial_{\beta} \psi^{b} -
\partial_{\mu} \psi^{a}  \partial_{\nu} \psi^{b} \bigg) \nonumber\\ &&  +
\frac{k \kappa_{5}^{2}}{2}  \left[ A_{2}^{2}  T_{\mu \nu}^{2} +
A_{1}^{2} T_{\mu \nu}^{1} \right] - \frac{3}{4} g_{\mu \nu} V,
\label{eq5: EIN-EFF1}
\end{eqnarray}
and the moduli fields' equations are:
\begin{eqnarray}
\Box \psi^{1} &=& - g^{\mu \nu} \gamma^{1}_{b c} (\partial_{\mu}
\psi^{b}) (\partial_{\nu} \psi^{c})
+ \frac{1}{2} \gamma^{1 c} \frac{\partial V}{\partial \psi^{c}}\nonumber\\
&& - \frac{k \kappa_{5}^{2}}{6} \frac{\partial U_{B}}{\partial
\psi^{1}} A_{1}^{2} \, g^{\mu \nu} T^{\,1}_{\mu \nu}, \label{eq5: SCA1-EFF2} \\
\Box \psi^{2} &=& - g^{\mu \nu} \gamma^{2}_{b c} (\partial_{\mu}
\psi^{b}) (\partial_{\nu} \psi^{c})
+ \frac{1}{2} \gamma^{2 c} \frac{\partial V}{\partial \psi^{c}}\nonumber\\
&& + \frac{k \kappa_{5}^{2}}{6} \frac{\partial U_{B}}{\partial
\psi^{2}} A_{2}^{2} \, g^{\mu \nu} T^{\,2}_{\mu \nu}. \label{eq5:
SCA2-EFF2}
\end{eqnarray}
Additionally, the matter conservation relations now read:
\begin{eqnarray}
\nabla^{\nu} T^{1}_{\mu \nu} &=& \frac{1}{A_{1}} \big[ (
\nabla_{\mu} A_{1})  g^{\alpha \beta} T^{1}_{\alpha
\beta} - 2 (\nabla^{\nu} A_{1}) T^{1}_{\mu \nu} \big], \\
\nabla^{\nu} T^{2}_{\mu \nu} &=& \frac{1}{A_{2}} \big[ (
\nabla_{\mu} A_{2}) g^{\alpha \beta} T^{2}_{\alpha \beta}
 - 2 (\nabla^{\nu} A_{2}) T^{2}_{\mu \nu} \big]. \quad \quad
\end{eqnarray}

In the next sections, we are going to study this theory in more
detail, analyzing a few examples (including the Randall-Sundrum
case) and studying the cosmological evolution of the moduli for
late cosmological eras. We shall pay special attention to the
observational constraints that can be imposed on this model.

\subsection{A Few Examples}

In this subsection we shall briefly analyze two well known cases,
namely, the exponential case $U_{B}(\psi) = V_{0} e^{\alpha
\psi}$, and the Randall-Sundrum case, which can be derived as
particular cases of the formalism described above.

\subsubsection{Exponential Case}

This case is particularly simple, and has been studied in detail
in several previous works \cite{Davis & Brax, Csaki etal, Youm,
Flanagan etal 2, DeWolfe etal, Br-CvdB-ACD-Rh, Kobayashi-Koyama}.
The potential to consider is $U_{B}(\psi) = V_{0} e^{\alpha
\psi}$, with $\alpha$ being a constant parameter of the theory.
Here, it is possible (and convenient) to parameterize the theory
using a new set of fields. Let us assume that $V_{0} > 0$ and
consider the following transformations:
\begin{eqnarray}
e^{2 \Phi_{1}} \cosh^{2} \Phi_{2} &=&  \frac{2}{1 + 2
\alpha^{2}}
e^{- (1 + 2 \alpha^{2}) \psi^{1} /2 \alpha} , \nonumber \\
e^{2 \Phi_{1}} \sinh^{2} \Phi_{2} &=&  \frac{2}{1 + 2
\alpha^{2}} e^{- (1 + 2 \alpha^{2}) \psi^{2} /2 \alpha} .
\end{eqnarray}
If $V_{0} < 0$ then we should interchange $\psi^{1}$ and $\psi^{2}$
in the previous definition of $\Phi_{1}$ and $\Phi_{2}$.
Inserting this notation back in the theory, it is
possible to obtain the following effective action:
\begin{eqnarray}
S &=& \frac{1}{k \kappa_{5}^{2}} \int d^{4}x \sqrt{- g} \bigg[ R -
\frac{12 \alpha^{2}}{1 + 2 \alpha^{2}}(\partial \Phi_{1})^{2}
\nonumber\\ &&
- \frac{6}{1 + 2 \alpha^{2}} (\partial \Phi_{2})^{2} - \frac{3}{4} V \bigg]  \nonumber\\
&& + S_{1}[\Psi_{1}, A_{1}^{2} g_{\mu \nu}] + S_{2}[\Psi_{2},
A_{2}^{2} g_{\mu \nu}],
\end{eqnarray}
where now the coefficients $A_{1}$ and $A_{2}$ are functions of
the fields $\Phi_{1}$ and $\Phi_{2}$, given by:
\begin{eqnarray}
A_{1}^{2} &=& e^{-2 \Phi_{1}} \frac{V_{0}}{k} \! \left[ \frac{1}{2} (1 + 2
\alpha^{2})
e^{2 \Phi_{1}} \cosh^{2} \Phi_{2} \right]^{\frac{1}{1 + 2 \alpha^{2}}}, \nonumber\\
A_{2}^{2} &=& e^{-2 \Phi_{1}} \frac{V_{0}}{k} \! \left[ \frac{1}{2} (1 + 2
\alpha^{2}) e^{2 \Phi_{1}} \sinh^{2} \Phi_{2} \right]^{\frac{1}{1
+ 2 \alpha^{2}}}. \qquad
\end{eqnarray}
This form of the theory was already obtained in
\cite{Br-CvdB-ACD-Rh}, in the moduli-space approximation approach,
and extensive analysis to it have been made.

\subsubsection{Randall Sundrum Case}

Finally, it is also instructive to examine the equations for the
Randall-Sundrum case. This can be obtained as a limiting case of
the previous example, by letting $\alpha \rightarrow 0$. The only
field surviving the limiting process is $\Phi_{2}$, and the
effective action is:
\begin{eqnarray}
S &=& \frac{1}{k \kappa_{5}^{2}} \int d^{4}x \sqrt{- g} \bigg[ R
- 6 (\partial \Phi_{2})^{2} - \frac{3}{4} V \bigg]  \nonumber\\
&& + S_{1}[\Psi_{1}, A_{1}^{2} g_{\mu \nu}] + S_{2}[\Psi_{2},
A_{2}^{2} g_{\mu \nu}],
\end{eqnarray}
where now the coefficients $A_{1}$ and $A_{2}$ are functions of
the field $\Phi_{2}$ alone, and are given by:
\begin{eqnarray}
A_{1}^{2} &=& \frac{V_{0}}{2 k} \cosh^{2} \Phi_{2}, \nonumber\\
A_{2}^{2} &=& \frac{V_{0}}{2 k} \sinh^{2} \Phi_{2}. \qquad
\end{eqnarray}
In this case, however, the field is not directly coupled to
matter.


\section{Late-Time Cosmology} \label{sec: Cosmology}

A simple and important application of the formalism developed above
is the study of cosmological solutions for the most recent epoch
of the Universe. In simple terms, current observations reveal that
the moduli fields are not relevant for the present evolution of
the universe and their dynamics is strongly suppressed. In this
section we shall analyze the possibility of stabilizing the moduli
fields $\psi^{1}$ and $\psi^{2}$. Interestingly, it will be found
that, in order to stabilize the system, constraints must be placed
on the global configurations of branes, particularly, on the class
of possible potentials and also on the position of the branes with
respect to the background.

\subsection{Cosmological Equations}

Let us derive the cosmological equations of motion of the
present system, valid for a flat, homogeneous and isotropic Universe.
We can derive the equations using the flat Friedmann-Robertson-Walker (FRW)
metric:
\begin{eqnarray}
ds^{2} = g_{\mu \nu} dx^{\mu} dx^{\nu}
= - d t^{2} + a^{2}(t) \delta_{i j} dx^{i} dx^{j}. \label{eq6:
ds-cos}
\end{eqnarray}
Here, $g_{\mu \nu}$ corresponds to the Einstein's frame metric,
conformally related to the physical metric, $\tilde g_{\mu \nu}$,
of the brane frame. Using the metric (\ref{eq6: ds-cos}), jointly
with the effective action (\ref{eq5: EFF Action 3}), we are left
with the following Friedmann equations:
\begin{eqnarray}
\frac{\ddot a}{a}  &=& - \frac{k \kappa_{5}}{12} A_{1}^{4}
(\rho_{1} + 3 p_{1}) - \frac{k \kappa_{5}}{12} A_{2}^{4} (\rho_{2}
+ 3 p_{2}) \nonumber\\ && - \frac{1}{4}
\gamma_{a b} \dot \psi^{a} \dot \psi^{b} + \frac{1}{4} V, \\
H^{2} &=& \frac{k \kappa_{5}}{6} A_{1}^{4} \rho_{1}  + \frac{k
\kappa_{5}}{6} A_{2}^{4} \rho_{2}  + \frac{1}{8} \gamma_{a b} \dot
\psi^{a} \dot \psi^{b} \nonumber\\ && + \frac{1}{4} V,
\end{eqnarray}
where $V$ is given in equation (\ref{eq5: Pot V}) and $H = \dot a
/ a$ is the Hubble parameter. The moduli fields equations, on the
other hand, are given by:
\begin{eqnarray}
\ddot \psi^{1} + 3 H \dot \psi^{1} &=& - \gamma_{a b}^{1} \dot
\psi^{a} \dot \psi^{b} - \frac{1}{2} \gamma^{1 c} \frac{\partial
V}{\partial \psi^{c}} \nonumber\\ && - \frac{1}{6} k
\kappa_{5}^{2}
\frac{\partial U_{B}}{\partial \psi^{1}} A_{1}^{4} (\rho_{1} - 3 p_{1}), \\
\ddot \psi^{2} + 3 H \dot \psi^{2} &=& - \gamma_{a b}^{2} \dot
\psi^{a} \dot \psi^{b} - \frac{1}{2} \gamma^{2 c} \frac{\partial
V}{\partial \psi^{c}} \nonumber\\ && + \frac{1}{6} k
\kappa_{5}^{2} \frac{\partial U_{B}}{\partial \psi^{2}} A_{2}^{4}
(\rho_{2} - 3 p_{2}).
\end{eqnarray}
Additionally, the matter conservation relations can be read as:
\begin{eqnarray}
\dot \rho_{1} + 3H (\rho_{1} + p_{1}) &=& - 3 \frac{\dot
A_{1}}{A_{1}} (\rho_{1} + p_{1}), \\
\dot \rho_{2} + 3H (\rho_{2} + p_{2}) &=& - 3 \frac{\dot
A_{2}}{A_{2}} (\rho_{2} + p_{2}).
\end{eqnarray}

The above set of equations are the corresponding equations of
motion describing the cosmological behaviour of the brane system. Since
the sigma-model metric is positive definite, the moduli fields
are not going to result in an accelerating universe.
Thus, in the present context, the only way to obtain an
accelerating universe is to consider supersymmetry breaking potentials.
A more detailed analysis of this system, taking into account the full
dependence on the extra kinetic terms coming from the sigma model
formulation is complicated, and shall be omitted in the
present work. However, since we are interested mostly in the low
energy regime of branes, we shall study the case in which the
moduli are slowly evolving compared to the Hubble parameter. We
examine this in the next subsection.

\subsection{Late-Time Cosmology}

Let us consider the case in which the moduli fields are slowly
evolving compared with the Hubble parameter, that is $|\dot
\psi^{1}|, |\dot \psi^{2}| \ll H$, and that the supersymmetry
breaking potentials are such that $u, v_{1}, v_{2} = 0$. Then, the
Friedmann equations of the system are:
\begin{eqnarray}
\frac{\ddot a}{a}  &=& - \frac{k \kappa_{5}}{12} A_{1}^{4}
(\rho_{1}
+ 3 p_{1}) - \frac{k \kappa_{5}}{12} A_{2}^{4} (\rho_{2} + 3 p_{2}), \quad \, \\
H^{2} &=& \frac{k \kappa_{5}}{6} A_{1}^{4} \rho_{1}  + \frac{k
\kappa_{5}}{6} A_{2}^{4} \rho_{2}.
\end{eqnarray}
The moduli fields equations, on the other hand, are given by:
\begin{eqnarray}
\ddot \psi^{1} + 3 H \dot \psi^{1} &=& - \frac{1}{6} k
\kappa_{5}^{2}
\frac{\partial U_{B}}{\partial \psi^{1}} A_{1}^{4} (\rho_{1} - 3 p_{1}), \\
\ddot \psi^{2} + 3 H \dot \psi^{2} &=& + \frac{1}{6} k
\kappa_{5}^{2} \frac{\partial U_{B}}{\partial \psi^{2}} A_{2}^{4}
(\rho_{2} - 3 p_{2}).
\end{eqnarray}
And the matter conservation relations can now be written as:
\begin{eqnarray}
\dot \rho_{1} + 3H (\rho_{1} + p_{1}) &=& 0, \\
\dot \rho_{2} + 3H (\rho_{2} + p_{2}) &=& 0.
\end{eqnarray}

\begin{figure}[ht]
\begin{center}
\includegraphics[width=0.45\textwidth]{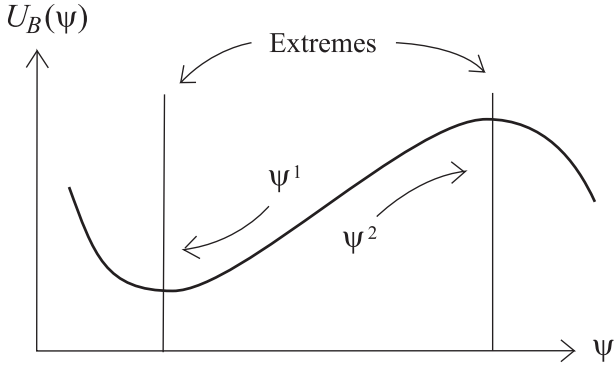}
\caption[Late Cosmology]{The figure shows the case when the branes
are matter dominated. The moduli fields $\psi^{1}$ and $\psi^{2}$
are driven to the minimum and maximum of the superpotential
$U_{B}$, respectively.} \label{F4}
\end{center}
\end{figure}

It is possible to appreciate that the moduli fields are driven by
the matter content of the branes, while the evolution of the
Universe remains unaffected by the moduli. In particular, if the
universe is matter dominated, then there is an attractor for the
moduli towards the extremes of the supersymmetric potential
$U_{B}$. To be more precise, $\psi^{1}$ will be attracted to the
minimum of $U_{B}$, while $\psi^{2}$ will be attracted to the
maximum of $U_{B}$. This is sketched in Figure \ref{F4}.
It is therefore sensible to assume that the present state of the
Universe is very close to this configuration, where the potential
is being extremize by the moduli. In the next subsection we assume
that this is the case in order to place observational constraints
on the model.

\subsection{Observational Constraints}

We now turn to the task of placing observational bounds on the
theory. In the last subsection we observed that the branes are
cosmologically driven by the matter content in them. In
particular, the evolution of the moduli is such that they
extremize the value of the supersymmetric potential $U_{B}$.
Therefore we expect that at late cosmological eras, such as the
present era, the branes are far apart and their tensions nearly at
their extreme values. This corresponds to an interesting global
condition on the configuration of the system, however, it can be
constrained by current observations. More precisely, the
measurement of the Post Newtonian Eddington coefficient $\gamma$
is constrained by the Very Long Baseline Interferometry
measurements of the deflection of radio waves by the Sun to be $
3.8\ 10^{-4} > 1 - \gamma  > - 2.6\ 10^{-4}$ at $68 \%$ confidence
level \cite{Eubanks, Will}. The parameter $\gamma$ is defined as:
\begin{eqnarray}
1 - \gamma = 2 \frac{\eta^{2}}{1 + \eta^{2}} \simeq 2 \eta^{2},
\end{eqnarray}
where $\eta^{2} = \gamma^{a b} \, \eta_{a} \eta_{b}$, and
$\eta_{a}$ is defined as:
\begin{eqnarray}
\eta_{a} = \frac{1}{A_{1}} \frac{\partial A_{1}}{\partial
\psi^{a}}.
\end{eqnarray}
Observe that $\eta_{a}$ is constructed from $A_{1}$, which is the
factor that appears in the action for the matter content at the
$\sigma_{1}$ brane. This is because we are assuming, without loss of
generality, that we are performing measurements from the
$\sigma_{1}$ brane. The quantity $\eta^{2}$ can be computed
exactly:
\begin{eqnarray}
\eta^{2} = \frac{1}{8} \left[ 1  - \frac{1}{2 k} U_{B}(\psi^{1})
A_{1}^{-2} \right].
\end{eqnarray}
Then, we finally obtain:
\begin{eqnarray}
1 - \gamma \simeq  \frac{2}{3} \left[ 1  - \frac{1}{2 k}
U_{B}(\psi^{1}) A_{1}^{-2} \right]. \label{eq6: gamma-edd}
\end{eqnarray}
(Note that this result is independent of $k$).
Recall that $A_{1}$ is related to an integral over the entire
bulk. This is a very important result: observational measurements
constrain the global configuration of the brane system. For
example, we can compute the $\gamma$ parameter for the exponential
potential. In this case, we obtain:
\begin{eqnarray}
1 - \gamma & \simeq & \frac{4}{3} \frac{\alpha^{2}}{1 + 2
\alpha^{2}} \nonumber \\ && + \frac{2}{3} \frac{1}{1 + 2
\alpha^{2}} e^{(1 + 2 \alpha^{2}) (\psi^{1} - \psi^{2})/2 \alpha},
\label{eq6: bound-exp}
\end{eqnarray}
which agrees with earlier computations \cite{Br-CvdB-ACD-Rh}. If
the branes are such that the moduli are extremizing the potential
$U_{B}$, then we have $\alpha ( \psi^{2} - \psi^{1}) \rightarrow +
\infty$, and the second term will disappear in (\ref{eq6:
bound-exp}). Then, we end up with a constraint on $\alpha$,
the parameter of the model:
\begin{eqnarray}
1 - \gamma & \simeq & \frac{4}{3} \frac{\alpha^{2}}{1 + 2
\alpha^{2}} ,
\end{eqnarray}
which gives $|\alpha| < 10^{-2}$. Interestingly, this bound does
not affect the global configuration of the brane, but only the parameter
of the theory. This is certainly a
problem since theoretically viable values for
$\alpha$ are usually of the order of unity. In the Randall-Sundrum
case there is no potential, and the moduli can not be
driven to any stable configuration. In this case, the constraint takes the
form:
\begin{eqnarray}
1 - \gamma & \simeq &  \frac{2}{3} \tanh^{2} \Phi_{2}. \label{eq6:
bound-RS}
\end{eqnarray}
Thus we see that in the Randall-Sundrum case a stabilization
mechanism for the radion field is necessary in order to agree with
observations.

\section{Conclusions} \label{sec: Conclusions}


In this paper we have studied several aspects of the low energy
regime of BPS brane-world models. We have developed a systematic
procedure to obtain solutions to the full system of equations,
consisting in a linear expansion of the fields about the static
vacuum solution of the system. As a result, it was shown that five
dimensional solutions can be obtained at any desired order in the
expansion. In particular, the projected Weyl tensor $E_{\mu \nu}$
and the loss parameter $\Delta \psi$ can be computed with any
desired accuracy. Additionally, and probably more important, an
effective 4-dimensional system of equations have been obtained.
Concordant with the method, these 4-dimensional effective
equations must be considered up to the desired order in the
expansion. For instance, we have analyzed in detail the zeroth
order effective theory, which agrees with the moduli-space
approximation \cite{prep}. At this order, the metrics of both
branes are conformally related, and the complete theory
corresponds to a bi-scalar tensor theory of gravity [equation
(\ref{eq5: EFF Action 3})]. The approach followed in this paper to
obtain the effective theory is similar to the one followed in a
previous work \cite{Kobayashi-Koyama}, where the cosmological
setup of the case $U_{B} \propto e^{\alpha \psi}$ was
investigated. However, the treatment in this article was more
general in two senses: we did not restrict the form of the
BPS-potential, and we deduced the complete Einstein's equations
governing the system.

We have also seen that the moduli fields --as defined in
the zeroth order part of the theory-- can be stabilized with the
help of effective couplings between the moduli and matter, that
arise naturally. This result comes from the fact that the
equations of motion for the projected bulk scalar field on the brane
depend only on the energy-momentum of the respective
brane. This allows satisfactory late cosmological configurations
and enables us to place observational constraints on the model.
An important result was the computation of the Post Newtonian Eddington
coefficient $\gamma$ in terms of the moduli [equation (\ref{eq6: gamma-edd})].
This result is relevant since $\gamma$ is currently the most
constrained parameter of General Relativity. For example, in the
exponential case it was found that $|\alpha| < 10^{-2}$.

Our results are applicable to many other
aspects of brane-worlds not considered in this paper. For example,
the development of the theory at the first order in the perturbed
metric would allow an ideal background to study the cosmological perturbations
and the CMB predictions \cite{Koyama, Rhodes etal}. Additionally, many results where the
exponential case $U_{B} \propto e^{\alpha \psi}$ was considered
can now be extended to the general case of an arbitrary potential.

\section*{Acknowledgements}

We are grateful to Philippe Brax and Carsten van de Bruck for
useful discussions. This work is supported in part by PPARC and
MIDEPLAN (GAP).

\appendix

\section{Definition of $X$, $Y_{\mu \nu}$
and $Z$} \label{app: Def}

Here we define the zeroth order and the linear expansions of
$\mathcal{X}$, $\mathcal{Y}_{\mu \nu}$ and $\mathcal{Z}$ of
equations (\ref{eq4: X-Y-Z}). The zeroth order quantities are:
\begin{eqnarray}
\lefteqn{ X_{0} = (\partial \psi_{0})^{2} - \frac{4}{3} \tilde R + 8 \,
\omega_{0}^{-1} \tilde \Box \omega_{0} + \omega_{0}^{2} u, } \\
\lefteqn{ Y^{0}_{\mu \nu} = \tilde G_{\mu \nu} + \frac{3}{4} \bigg[
\frac{1}{2} \tilde g_{\mu \nu} (\partial \psi_{0})^{2} -
\partial_{\mu} \psi_{0} \partial_{\nu} \psi_{0} \bigg] } \nonumber\\ && {}
 + (N_{0} \omega_{0}^{2})^{-1}
\Big[ \tilde g_{\mu \nu} \tilde \Box (N_{0} \omega_{0}^{2})  -
\tilde \nabla_{\mu} \tilde \nabla_{\nu} (N_{0} \omega_{0}^{2})
\nonumber\\ && {} - 3 \tilde g_{\mu \nu} \partial^{\alpha} \omega_{0}
\partial_{\alpha} (N_{0} \omega_{0}) - 3 \partial_{\mu} \omega_{0}
\partial_{\nu} (N_{0} \omega_{0}) \nonumber\\ && {} - 3 \partial_{\nu} \omega_{0}
\partial_{\mu} (N_{0} \omega_{0}) \Big] + \frac{3}{8} \omega_{0}^{2} \tilde g_{\mu \nu}
u, \\
\lefteqn{ Z_{0} = - \frac{1}{N_{0}} \tilde g^{\mu \nu} \tilde \nabla_{\mu}
(N_{0}
\partial_{\nu} \psi_{0})} \nonumber \\ && {} - 2 \omega_{0}^{-1}
\tilde g^{\rho \lambda}
\partial_{\lambda} \omega_{0} \partial_{\rho} \psi_{0} +
\frac{\omega_{0}^{2}}{2} \frac{\partial u }{\partial \psi_{0}} .
\end{eqnarray}
Meanwhile, the linear terms in the expansions of $\mathcal{X}$,
$\mathcal{Y}_{\mu \nu}$ and $\mathcal{Z}$ are given
by the following expressions:
\begin{eqnarray}
\lefteqn{ X = 2 \,
\partial \psi_{0}
\partial \varphi  - \frac{4}{3} (\tilde \nabla_{\mu}
\tilde \nabla_{\mu} h^{\mu \nu} - \tilde \Box h) }  \nonumber\\ && {} - 4 \,
\omega_{0}^{-1} \tilde g^{\alpha \beta} \tilde g^{\rho \lambda} (
\tilde \nabla_{\alpha} h_{\lambda \beta} + \tilde \nabla_{\beta} h_{\alpha
\lambda} \nonumber\\ && {} - \tilde \nabla_{\lambda} h_{\alpha \beta})
\tilde \nabla_{\rho}
\omega_{0} , \\
\lefteqn{ Y_{\mu \nu} =  \frac{1}{2} \big( \tilde \nabla_{\sigma}
\tilde \nabla_{\nu} h^{\sigma}_{\mu} + \tilde \nabla_{\sigma} \tilde
\nabla_{\mu}
h^{\sigma}_{\nu} - \tilde \nabla_{\mu} \tilde \nabla_{\nu} h
- \tilde \Box h_{\mu \nu}
} \nonumber\\ && {} - \tilde g_{\mu \nu} \tilde \nabla_{\alpha}
\tilde \nabla_{\beta}
h^{\alpha \beta} + \tilde g_{\mu \nu} \tilde \Box h \big)  + \omega_{0}^{-1}
\tilde g^{\rho \lambda} \big(
\tilde \nabla_{\mu} h_{\lambda \nu} \nonumber\\ &&
{} + \tilde \nabla_{\nu} h_{\mu \lambda} - \tilde \nabla_{\lambda} h_{\mu \nu}
 - 2 \tilde g_{\mu \nu} \tilde \nabla_{\alpha}
h_{\lambda}^{\alpha} + \tilde g_{\mu \nu}
\tilde \nabla_{\lambda} h \big) \tilde \nabla_{\rho} \omega_{0}  \nonumber\\
&& + \frac{3}{4} \big( \tilde g_{\mu \nu} \tilde g^{\alpha \beta}
\partial_{\alpha} \psi_{0}
\partial_{\beta} \varphi - \partial_{\mu} \psi_{0} \partial_{\nu} \varphi
- \partial_{\nu} \psi_{0} \partial_{\mu} \varphi \big)
\nonumber\\ && +  \tilde g_{\mu \nu} \tilde \Box \phi - \tilde \nabla_{\mu}
\tilde \nabla_{\nu} \phi + \tilde g_{\mu \nu} \omega_{0}^{-1}
\tilde g^{\alpha \beta}
\partial_{\alpha} \omega_{0} \partial_{\beta} \phi
\nonumber\\ && {} + \omega_{0}^{-1} [ \partial_{\mu} \omega_{0}
\partial_{\nu}
\phi + \partial_{\nu} \omega_{0} \partial_{\mu} \phi ] \nonumber\\
&& + \frac{1}{N_{0}} \left( 2 \tilde g_{\mu \nu} \tilde g^{\alpha \beta}
\partial_{\alpha} N_{0} \partial_{\beta} \phi
- \partial_{\mu} N_{0} \partial_{\nu} \phi - \partial_{\nu} N_{0}
\partial_{\mu} \phi \right) \nonumber\\ &&
+ \frac{1}{2 N_{0}} \big( \tilde \nabla_{\mu} h_{\lambda \nu} +
\tilde \nabla_{\nu} h_{\mu \lambda} - \tilde \nabla_{\lambda} h_{\mu \nu}
\nonumber\\ && {} - 2 \tilde g_{\mu \nu} \tilde g^{\alpha \beta}
\tilde \nabla_{\alpha}
h_{\beta \lambda} + \tilde g_{\mu \nu} \tilde \nabla_{\lambda} h \big)
\tilde g^{\rho \lambda}
\tilde \nabla_{\rho} N_{0} , \\
\lefteqn{ Z =  \frac{1}{2} \tilde g^{\rho \lambda} \tilde g^{\mu \nu} (
\tilde \nabla_{\mu} h_{\lambda \nu} + \tilde \nabla_{\nu} h_{\mu \lambda}
- \tilde \nabla_{\lambda} h_{\mu \nu}) \partial_{\rho} \psi_{0}} \nonumber\\
&& {} -
\tilde g^{\mu \nu} \tilde \nabla_{\mu} \phi \tilde \nabla_{\nu}
\psi_{0}  - \tilde \Box \varphi - 2 \omega_{0}^{-1} \tilde g^{\rho \lambda}
\partial_{\lambda} \omega_{0} \partial_{\rho} \varphi .
\end{eqnarray}
To compute the $i$-th order terms of the linear expansions, with
$i>0$, it is enough to add an index ``$i$'' to every linear
variable in the expressions above.

\section{Computation of the Zeroth Order Theory} \label{app: Comp}

Here we indicate how to compute the integral in the left hand side of
equation (\ref{eq5: eff-0-1}). The integral present in equation
(\ref{eq5: eff-0-2}) can be solved in a similar way. First,
it is important to note the following identity:
\begin{eqnarray}
\partial_{\mu} (N_{0} \omega_{0}) = -N_{0} \alpha_{0} \omega_{0}
\partial_{\mu} \psi_{0} - \frac{4}{U_{B}} ( \partial_{\mu}
\omega_{0})'.
\end{eqnarray}
Additionally, recall that we can parameterize the $z$ coordinate using
the monotonic zeroth order solution for the scalar field $\psi_{0}(z)$.
That is, we can write:
\begin{eqnarray}
dz = \frac{1}{N_{0}} \left( \frac{\partial U_{B}}{\partial \psi_{0}} \right)^{-1} \!\!\! d \psi_{0}.
\end{eqnarray}
Using the previous equations, with boundaries
$\psi_{0}(z_{1}) = \psi^{1}$ and $\psi_{0}(z_{2}) = \psi^{2}$,
then the mentioned integral can be solved as:
\begin{eqnarray}
k \!\int_{z_{1}}^{z_{2}} \!\!\!\! dz N_{0} \omega_{0}^{2} Y_{\mu
\nu}^{0} &=& \Omega^{2} \tilde G_{\mu \nu} + \tilde g_{\mu \nu}
\tilde \Box \Omega^{2} - \tilde \nabla_{\mu} \tilde \nabla_{\nu}
\Omega^{2} \nonumber\\ && + \tilde g_{\mu \nu} \frac{3 k}{8}
\int^{\psi^{2}}_{\psi^{1}} \!\!\!\!\! d\psi \left( \frac{\partial
U_{B}}{\partial \psi} \right)^{-1} \!\!\! \omega^{4} u \nonumber\\
&& + \frac{3}{4} \gamma_{a b} \bigg( \frac{1}{2} \tilde g_{\mu
\nu}  \tilde g^{\alpha \beta}
\partial_{\alpha} \psi^{a}  \partial_{\beta} \psi^{b} \nonumber\\ && -
\partial_{\mu} \psi^{a}  \partial_{\nu} \psi^{b} \bigg)  ,
\end{eqnarray}
where $\Omega^{2}$ and $\gamma_{a b}$ are defined as in Section
\ref{sec: LER-eff}. To finish, we should mention that in obtaining this result it was
useful to notice that $\Omega^{2}$ does not only depend on
$\psi^{1}$ and $\psi^{2}$ through the integration limits in
(\ref{eq5: Big-Omega}), but also through $\omega^{2}$ present in
the integrand. Recall that $\omega$ is normalized to be $1$ at the
position of the first brane, and therefore depends on $\psi^{1}$.
This in turn means that the space-time derivative of $\Omega^{2}$
will have the form:
\begin{eqnarray}
\partial_{\mu} \Omega^{2} &=& \alpha_{1}^{-1}\left[ \frac{1}{2} \Omega^{2}
- \frac{k}{U_{B}(\psi^{1})}   \right] \partial_{\mu} \psi^{1} \nonumber\\
&& + \alpha_{2}^{-1} \frac{\bar \omega^{2} k}{U_{B}(\psi^{2})}
\partial_{\mu} \psi^{2}.
\end{eqnarray}

\end{document}